\documentclass[useAMS,usenatbib]{mn2e}
\usepackage{times}
\usepackage{amssymb}
\newcommand{\kms}{km~s$^{-1}$}
\usepackage{graphics,epsfig}
\newcommand{\simgeq}{\ensuremath{\widetilde{>}}}
\newcommand{\simleq}{\ensuremath{\widetilde{<}}}

\title[FIGGS: Faint Irregular Galaxies GMRT Survey]{FIGGS: Faint Irregular Galaxies GMRT Survey $-$ Overview, observations and first results}
\author[Begum et al.]
{
Ayesha Begum$^{1}$\thanks{E-mail:ayesha@ast.cam.ac.uk},
Jayaram N. Chengalur$^{2}$,
I. D. Karachentsev$^{3}$, 
M. E. Sharina$^{3}$  
\newauthor and S. S. Kaisin$^{3}$
\\
\\
$^{1}$Institute of Astronomy, University of Cambridge, Madingley Road, Cambridge, CB3 0HA, UK\\
$^{2}$National Centre for Radio Astrophysics, Post Bag 3, Ganeshkhind, Pune 411 007, India\\
$^{3}$Special Astrophysical Observatory, Nizhnii Arkhys 369167, Russia\\
}
\begin{document}

\date{}

%\pagerange{\pageref{firstpage}--\pageref{lastpage}} \pubyear{}

\maketitle

\label{firstpage}

\begin{abstract}

 The Faint Irregular Galaxies GMRT Survey (FIGGS) is a Giant Metrewave Radio Telescope (GMRT)
based HI imaging survey of a systematically selected sample of extremely faint nearby dwarf 
irregular galaxies. The primary goal of FIGGS is to provide a comprehensive and statistically
robust characterization of the neutral inter-stellar medium properties of faint, 
gas rich dwarf galaxies.  The FIGGS galaxies represent the extremely low-mass
end of the dwarf irregular galaxies population, with a median M${\rm{_B\sim-13.0}}$ and median
HI mass of $\sim 3 \times 10^7$ M$_\odot$, extending the baseline in mass and luminosity
space for a comparative study of galaxy properties. The HI data is supplemented with observations
at other wavelengths. In addition, distances accurate to $\sim$ 10\% are available for most
of the sample galaxies. This paper gives an introduction to FIGGS, describe the GMRT observations and presents the first 
results from the HI observations. From the FIGGS data we confirm the trend of increasing HI to optical diameter
ratio with decreasing optical luminosity; the median ratio of D$_{\rm HI}$/D$_{\rm Ho}$ for 
the FIGGS sample is 2.4. Further, on comparing our data with aperture synthesis surveys of bright spirals, 
we find  at best marginal evidence for a decrease in average surface density with decreasing
HI mass. To a  good approximation the disks of gas rich galaxies, ranging over 3 orders of magnitude
in HI mass, can be described as being  drawn from a family with constant HI surface density. 
\end{abstract}

\begin{keywords}
          galaxies: dwarf --
          galaxies: kinematics and dynamics --
          radio lines: galaxies
\end{keywords}

\section{Introduction}
\label{sec:intro}

     HI 21cm aperture synthesis observations of nearby spiral galaxies is a
mature field with over three decades of history -- probably something of the order
of a thousand galaxies have already been imaged. However observers have tended 
to focus on bright ($\sim$ L$_*$) galaxies with HI masses $\gtrsim 10^9$ M$_\odot$. 
HI observations of faint dwarf galaxies (M${\rm{_B \gtrsim -17}}$) generally 
require comparatively long integration times, and such galaxies have hence not 
been studied in similar numbers. While there have been some systematic
HI surveys of dwarf galaxies  (\citealp{swater99,stil02}), these have generally 
been restricted to the brighter  (M${\rm{_B \simleq -14}}$) dwarfs. 

       In hierarchical models of galaxy formation, nearby dwarf galaxies would, 
in some ways, be analogs of the primordial building blocks of large galaxies.
A systematic HI survey of the faintest dwarf galaxies could provide data that
would be useful for a diverse range of studies, ranging from, for example, 
testing the predictions of cold dark matter models (e.g \cite{simon07,blanton07}), checking
if such systems could be the  host population of quasar absorption line systems 
(e.g. \cite{zwaan05,kc05}) etc. As  the most chemically unevolved systems in the present-day 
galaxy population, the faintest dwarfs provide unique laboratories for understanding 
star formation and galaxy evolution in extreme environments, i.e. low 
metallicity, low dust content, low pressure, low shear, and low escape velocity
(e.g. \cite{ekta06}).

In this paper we describe and present the first results from a Giant Meterwave
Radio Telescope  (GMRT) based HI imaging study of faint dwarf galaxies $-$ the Faint 
Irregular Galaxies GMRT Survey (FIGGS). The primary  goal of FIGGS is to obtain high quality
observations of the atomic ISM  for a large, systematically selected sample 
of faint, gas rich,  dwarf irregular (dIrr) galaxies. Our GMRT HI images are supplemented 
by single dish HI observations,  HST  V and I band images and ground based H$\alpha$ images
from the 6-m BTA telescope.  Additionally,  the HII region abundances and H$\alpha$ rotation
curves are being obtained on the William Herschel Telescope (WHT), Isaac Newton Telescope (INT) 
telescopes on La Palma and 6-m Russian BTA telescope, respectively.

This paper is organised as follows. In Section~\ref{sec:sample} we describe the design and
the properties of the galaxy sample. The main science drivers for FIGGS are described in
Section.\ref{sec:science}. The GMRT observations are described in Section~\ref{sec:obs} and
the results of the survey are presented and discussed in Section~\ref{sec:result}.

\begin{table*}
\begin{center}
\label{tab:figgs_sample}
\begin{center}
\caption{ The FIGGS sample}
\end{center}
\begin{tabular}{|ccccccccccc|}
\hline
Galaxy & $\alpha$ (J2000) & $\delta$ (J2000) &M${\rm{_{B}}}$ &D$_{\rm{Ho}}$ & B-V &Dist& D estm&Group&i$_{opt}$ & Ref\\
       &(h~m~s)& ($^\circ$~$^\prime$~$^{\prime\prime}$) &(mag)  &($^\prime$)&(mag) &(Mpc)     &&&(deg) &\\
\hline\hline
SC 24& 00 36 38.00 & $-$ 32 34 28& $-$8.39& 0.6& $-$& 1.66& Sculptor grp(?) &distant Irr  & 57 &\\
And IV & 00 42 32.30 & +40 34 19 & $-$12.23 & 1.1 & 0.47 & 6.3 &rgb & Field &  41 & 18,19\\
DDO 226    &      00 43 03.80   &$-$22 15 01 &  $-$14.17       &3.2&0.4 &4.9& rgb&Sculptor& 72 &1\\
DDO 6      &      00 49 49.30 &$-$21 00 58&$-$12.5&2.1&0.32&3.34&rgb&Sculptor&69 & 1\\
UGC 685      &      01 07  22.30   &+16 41 02 & $-$14.31      & 2.2&0.52   &4.5&rgb&Field&46&1\\
KKH 6         &   01 34  51.60 &+52 5 30 &$-$12.42       &0.9 & 0.43  &3.73&rgb&IC~342/Maffei&55 &13,17\\
KK 14         &   01 44  42.70 &+27 17 16 & $-$12.13   &1.6 & 0.42 &7.2&N672 grp&N672&71&13\\
KKH 11        &   02 24  35.00 &+56 0 42 &$-$13.35  &1.7 & $-$  &3.0&Mafeii grp&IC~342/Maffei&59 &\\
KKH 12        &   02 27  27.00 &+57 29 16 &$-$13.03 &2.2   &$-$  &3.0&Maffei grp&IC~342/Maffei&78\\
KK 41 &      04 25  15.60 &+72 48 21&$-$14.06  &2.6   &  0.63 &3.9&rgb&IC~342&57 & 2\\
UGCA 92        &   04 32 00.30 &+63 36 50 & $-$15.65  &2.0  & 1.1 &3.01&rgb&IC~342/Maffei&62 &3, 17\\
KK 44&  04 53 06.90& +67 05 57& $-$11.85& 1.4& 0.58& 3.34& rgb& IC432 & 62 &4\\
KKH 34  &   05 59  41.20 &+73 25 39 & $-$12.30 &1.0   &  0.4&4.6&rgb&M81&58&13\\
E490-17   &      06 37  56.60 &$-$25 59 59& $-$14.46     & 2.0&0.51 &4.2&rgb&Field&42&5\\
%KKH 37,Mai16 &   06 47  45.80 &+80 7 26  & $-$12.60  &  1.2&   &3.39& rgb&M81&\\
HIZSS003     &   07 00  29.30  &$-$04 12 30  &$-$12.6   & 2.0 &$-$  &1.69&rgb&Field&55&\\
UGC 3755      &     07 13 51.80& +10 31 19&$-$14.90&1.8& 0.55 &6.96&rgb&Field&55& 6,15\\
DDO 43&     07 28  17.20 &+40 46 13 &$-$14.75    &1.8 &0.31 &7.8&rgb&Field&48 &7\\
KK 65       &     07 42  31.20 &+16 33 40 &$-$14.29 & 0.9 & 0.54 &7.62&rgb&Field& 58 &8\\
UGC 4115     &      07 57   01.80  &+14 23 27&  $-$14.27  &1.5 &  0.47&7.5&rgb&Field&58 &13\\
KDG 52& 	08 32 56.00 & +71 01 46& $-$11.49& 1.3& 0.24& 3.55& rgb& M81& 24 &4\\
UGC 4459& 08 34 06.50& +66 10 45&$-$13.37&1.6&0.45& 3.56&rgb&M81&30 &4\\
KK 69         &   08 52  50.70 &+33 47 52 & $-$12.76   & 2.0 &  $-$ &7.7& N2683 grp&N2638&42\\
UGC 5186        &   09 42  59.80 &+33 15 52 &$-$12.98   &1.3& 0.49 &6.9& h&Field&83&13\\
UGC 5209  &   09 45   04.20 &+32 14 18 &$-$13.15   &0.9 &  0.56 &6.7& h&Field&17& 8\\
UGC 5456     &      10 07  19.70  &+10 21 44&$-$15.08         &1.9&0.33 &5.6& rgb&Field&62& 9\\
HS~117        &   10 21  25.20 &+71 06 58& $-$11.83&  1.5 & $-$ &3.96&rgb&Field&55\\
UGC 6145 &   11 05  35.00  &$-$01 51 49  &   $-$13.14 &1.7$^*$ &$-$  &7.4&h&Field&64\\
UGC 6456     &      11 28   00.60 &+78 59 29&$-$14.03        &1.5$^*$&  0.38 &4.3& rgb&M81&60&11\\
UGC 6541        &   11 33 29.10   &+49 14 17 & $-$13.71       &1.6&  0.41 &3.9&rgb&CVn~I&57& 10\\
NGC 3741     &      11 36   06.40 &+45 17 7 &$-$13.13    & 1.7   &  0.37 &3.0&rgb&CVn~I&58& 7\\
KK 109        &   11 47 11.20 &+43 40 19 & $-$9.73 &  0.6 & 0.36  &4.5&rgb&CVn~I&49 &13\\
DDO 99&     11 50 53.00 &+38 52 50 & $-$13.52   &3.5&  0.38 &2.6&rgb&CVn~I&71&13\\
E379-07&   11 54  43.00 &$-$33 33 29& $-$12.31 & 1.1  & 0.23 &   5.2&rgb&Cen~A&44&13\\
KK 127        &   12 13  22.70 &+29 55 18 & $-$15.30 & 1.0 & 0.4   &13.0&ComaI grp&ComaI&69& 17\\
E321-014     &   12 13  49.60 &$-$38 13 53& $-$12.70  &1.4&  0.41 &3.2&rgb&Cen~A&67&13\\
UGC 7242        &   12 14 07.40&+66 05 32&$-$14.06&1.9& 0.4  &5.4&rgb&M81&68 &13,17\\
CGCG 269-049&    12 15 46.70& +52 23 15& $-$13.25&1.8& 0.4& 4.9& rgb& CVn I&77 &16\\
UGC 7298&     12 16 28.60 & +52 13 38& $-$12.27 & 1.1& 0.29& 4.21&rgb&CVn I& 58 &4\\
UGC 7505  &   12 25  17.90 &+26 42 53& $-$15.55  &1.0 &0.45  &12.8&tf&ComaI&84 & 6\\
KK 144        &   12 25  27.90 &+28 28 57 &$-$12.59 & 1.5 &    0.4 &6.3&h&CVn~I&74 &13\\
DDO 125 &   12 27  41.80 &+43 29 38& $-$14.16&4.2  &  0.59 &2.5&rgb&CVn~I& 58& 6\\
UGC 7605        &   12 28 39.00 &+35 43 05&$-$13.53&2.2&  0.41 &4.43&rgb&CVn~I&44& 13\\
UGC 8055        &   12 56 04.00 &+03 48 41 & $-$15.49&  1.4  & 0.24 &17.4&tf&Field&40& 9\\
GR 8         &   12 58 40.40 & +14 13 03& $-$12.11& 2.2& 0.32& 2.1& rgb& Field & 25 &4\\
UGC 8215        &   13 08   03.60  &+46 49 41 &$-$12.26    &1.0 & 0.38  &4.5&rgb&CVn~I&47 &12, 17\\
DDO 167 &   13 13  22.80 &+46 19 11& $-$12.70    & 1.6& 0.32    &4.2&rgb&CVn~I&58 &12\\
KK 195        &   13 21   08.20  &$-$31 31 45 &$-$11.76    & 1.3&  $-$ &5.22&rgb&M83&65\\
KK 200        &   13 24  36.00 &$-$30 58 20& $-$11.96       &1.3& 0.4  & 4.6&rgb&M83&53&13\\
UGC 8508        &   13 30 44.40 &+54 54 36&$-$12.98&2.0&  0.45 &2.6&rgb&CVn~I&55& 6\\
E444-78&   13 36  30.80 &$-$29 14 11 &$-$13.3      &1.2$^*$ &0.49  & 5.25& rgb&M83&68& 13\\
UGC 8638        &   13 39  19.40 &+24 46 33 &$-$13.68& 1.2&   0.51 &   4.27&rgb&CVn~I&49 &13, 17\\
DDO 181 &   13 39  53.80 &+40 44 21& $-$13.03 &1.6 &  0.46  &3.1&rgb&CVn~I& 57 &13\\
I4316&   13 40  18.10 &$-$28 53 40& $-$13.90    &  1.6& $-$   &4.4&rgb&M83&52\\
DDO 183       &   13 50 51.10 &+38 01 16&$-$13.17 &1.7&  0.31 &3.24&rgb&CVn~I&75 &9\\
UGC 8833        &   13 54  48.70 &+35 50 15 &  $-$12.42  & 1.3  & 0.42   &3.2&rgb&CVn~I&28 &13\\
KK 230       & 14 07 10.70 &+35 03 37&  $-$9.55& 1.7 & 0.4& 1.9& rgb& Field& 35 &4\\ 
DDO 187       &   14 15 56.50  &+23 03 19 &  $-$12.51  &   2.28 &  0.28 &2.5&rgb&Field&42&1\\
P51659       &   14 28   03.70 &$-$46 18 06 &$-$11.83 &2.4$^*$&   $-$ &3.6& rgb&Cen~A&71\\
KKR 25       &   16 13 47.60 & +54 22 16 & $-$9.96 & 1.2 & $-$ & 1.86& rgb& Field & 55 \\
KK 246        &   20 03  57.40  &$-$31 40 54  & $-$13.69&1.3  &   0.58  &7.83& rgb&Field& 68& 13, 17\\
%KK250 &          20 30 15.30 & +60 26 25& $-$16.14& 1.8& 0.91& 5.6 &N6946 grp& N6946& 62& 14\\
%KK251 &          20 30 32.60 & +60 21 13& $-$14.32& 1.6& 0.37& 5.6 &N6946 grp& N6946& 66 &14\\
%DDO 210      &  20 46 53.00 & $-$12 50 57& $-$11.09& 3.6& 0.24 & 1.0 &rgb & Local grp & 62 &4\\
%UA438        &   23 26  27.50 &$-$32 23 26 & $-$12.94    &   2.4&  0.4 &2.2&rgb&Sculptor&38&13\\
%KKH98        &   23 45  34.00 &+38 43 04 & $-$10.78 &1.1   &    0.21 &2.5&rgb&Field&58 &13\\
\hline
\hline
\end{tabular}
\end{center}
%{References:
{\vskip-0.2cm
{\hskip-11.0cm *: Optical diameter measured at 25.0 mag arcsec$^{-2}$.}}
%}
\end{table*}

\begin{table*}
\addtocounter{table}{-1}
\begin{center}
%\label{tab:figgs_sample}
\begin{center}
\caption[Continued]{({\it{continued}}) The FIGGS sample}
\end{center}
\begin{tabular}{|ccccccccccc|}
\hline
Galaxy & $\alpha$ (J2000) & $\delta$ (J2000) &M${\rm{_{B}}}$ &D$_{\rm{Ho}}$ & B-V &Dist& D estm&Group&i$_{opt}$ & Ref\\
       &(h~m~s)&($^\circ$~$^\prime$~$^{\prime\prime}$) &(mag)  &($^\prime$)&(mag) &(Mpc)     &&&(deg) &\\
\hline\hline
KK 250, UGC11583 &          20 30 15.30 & +60 26 25& $-$14.54& 1.8& 0.91& 5.6 &N6946 grp& N6946& 62& 14\\
KK 251 &          20 30 32.60 & +60 21 13& $-$13.72& 1.6& 0.37& 5.6 &N6946 grp& N6946& 66 &14\\
DDO 210      &  20 46 53.00 & $-$12 50 57& $-$11.09& 3.6& 0.24 & 1.0 &rgb & Field & 62 &4\\
UGCA 438        &   23 26  27.50 &$-$32 23 26 & $-$12.94    &   2.4&  0.42 &2.2&rgb&Sculptor&38&13\\
KKH 98        &   23 45  34.00 &+38 43 4 & $-$10.78 &1.1   &    0.21 &2.5&rgb&Field&58 &13\\
\hline
\end{tabular}
\end{center}
{References:
1-\cite{vanzee00}
2-\cite{kk99}
3-\cite{kk96}
4-\cite{b06}
5-\cite{parodi02}
6-\cite{makarova99}
7-\cite{taylor05}
8-\cite{makarova02}
9-\cite{hunter06}
10-\cite{bremnes00}
11-\cite{hopp95}
12-\cite{bremnes99}
13-Sharina et al. 2008 (in preparation)
14-\cite{bc04b}
15-\cite{tully06}
16-\cite{corbin08}
17-\cite{kk06}
18-\cite{and14}
19-Chengalur et al. 2008 (in preparation)
}
\end{table*}

\section{FIGGS: Sample definition and properties}
\label{sec:sample}

The Faint Irregular Galaxies GMRT Survey $-$ FIGGS, is a large observing program aimed at providing
a comprehensive and  statistically robust characterisation of the neutral ISM properties of
faint, gas rich, dwarf irregular galaxies using the Giant Metrewave Radio Telescope (GMRT).
The FIGGS sample forms a subsample of the Karachentsev et al.(2004) catalog
of galaxies within $\sim$ 10 Mpc.  Specifically, the FIGGS sample consists 
of  65 faint dwarf irregular (dIrr)galaxies with:

\begin{enumerate}
\item absolute blue magnitude, M${\rm{_B}} \simgeq -14.5$, 
\item HI flux integral $>$ 1 Jy kms$^{-1}$ 
\item optical B band major axis $\simgeq$ 1.0 arcmin. 
\end{enumerate}

The sample choice was dictated by a balance between achieving the scientific goals
described in Section.\ref{sec:science} and the practical limitations of the observing time.
We note that the above mentioned criterion on the optical B band major axis was not
strictly followed in few cases.  Some unusual, very faint dwarf galaxies, with
 optical B band major axis $<$ 1 arcmin were still included in our sample, as they are 
interesting cases to study in detail in HI.
Further, for some of the galaxies in the FIGGS sample, fresh estimates of the distance
(obtained after our observations were complete) imply absolute magnitudes slightly larger
than the cut off above. These galaxies have however been retained in the sample.
Some properties (mainly derived from optical observations) of galaxies in the FIGGS sample 
are listed in Table~\ref{tab:figgs_sample}. The columns are as follows: Column(1)~the galaxy name, 
Column(2)\&(3)~the equatorial coordinates (J2000), Column(4)~the absolute blue magnitude 
(corrected for galactic extinction), Column(5)~the Holmberg diameter in arcmin, Column(6)~the 
(B-V) colour, Column(7)~ the distance in Mpc, Column(8)~the method used to measure the 
distance $-$ from the tip of the red giant branch (rgb), from membership in a group with
known distance (grp), from the Tully-Fisher relation (tf), and from the Hubble flow (h). 
Column(9) gives the group membership of the galaxy, Column(10)~the inclination determined from 
optical photometry (and assuming an intrinsic thickness, q$_o$=0.2) and Column(11)~the reference 
for the (B-V) colour,  and/or revised distance. 
The data presented in the Table~\ref{tab:figgs_sample} (except for the colour) 
is taken from \cite{kk04} catalog, except that revised distances have been adopted, 
if available. As can be seen from the Table~\ref{tab:figgs_sample}, tip of the red giant 
branch (rgb) distances (which are generally accurate to $\sim$ 10\%) are available for most 
of the galaxies in our sample.

\begin{figure*}
\psfig{file=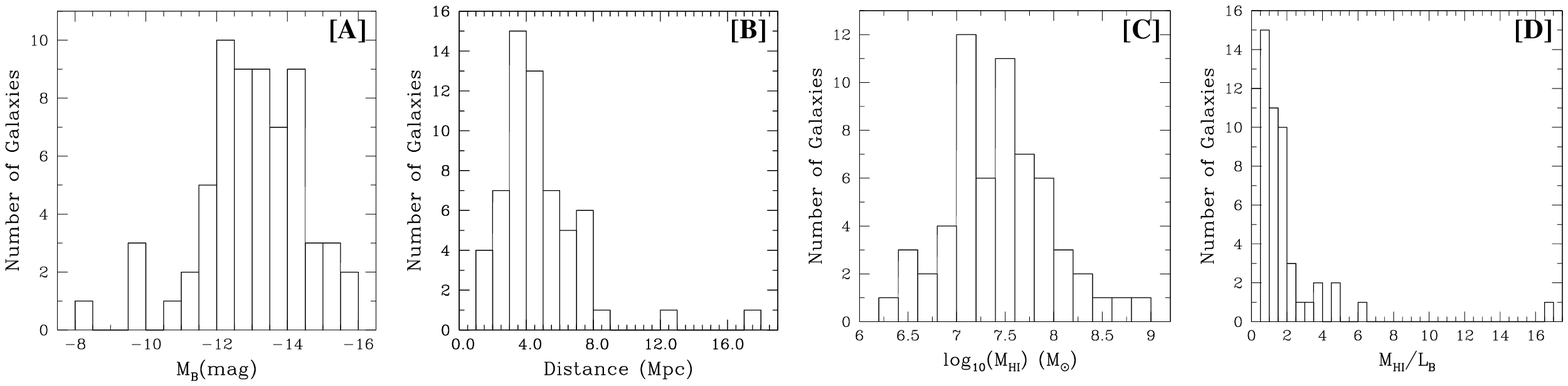,width=7.0truein}
\caption{ The histogram of M${\rm{_B}}$ (panel [A]), distance (panel [B]), logarithm
of the HI mass (panel [C]) and the HI mass to light ratio, M${\rm{_{HI}/L_B}}$ (panel [D])
for the FIGGS sample.
}
\label{fig:hist_figgs}
\end{figure*}

\begin{figure*}
\psfig{file=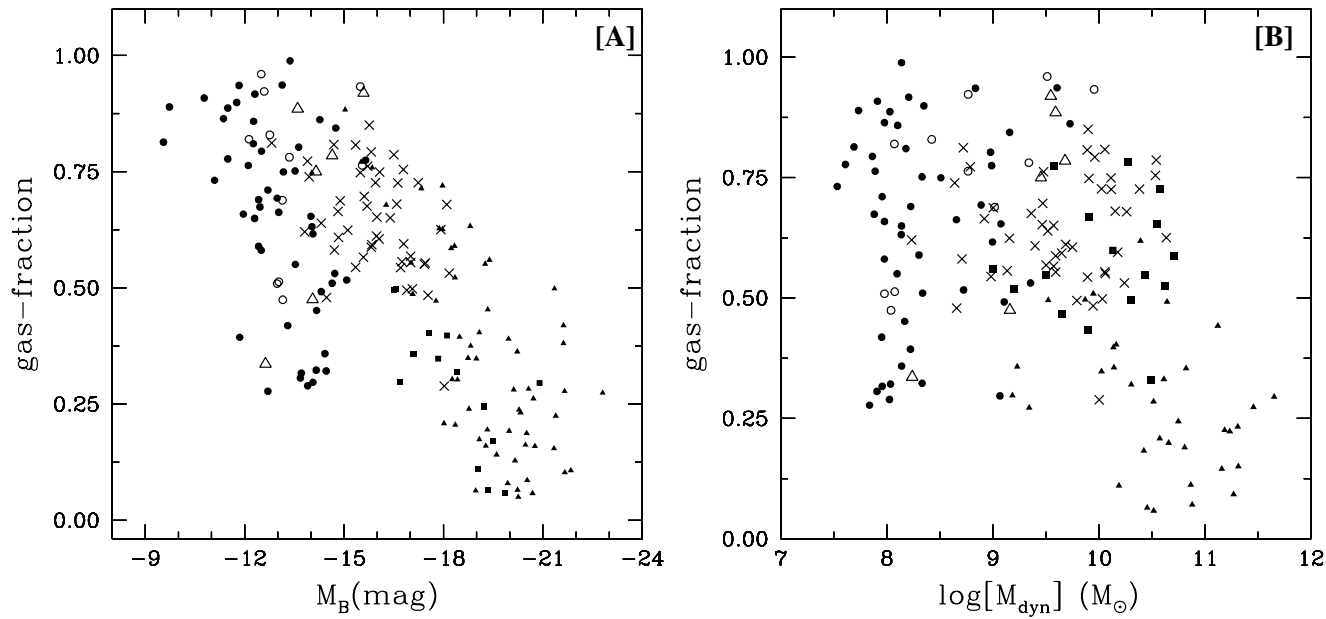,width=7.0truein}
\caption{
The gas fraction of FIGGS galaxies (circles) plotted as 
a function of the absolute blue magnitude (left) and dynamical mass (right).
FIGGS galaxies with TRGB distances are shown as solid circles, whereas
the remaining FIGGS galaxies are shown as empty circles.
The same quantity is also plotted for the galaxies in  literature with
interferometric HI maps.  The gas fraction (f$_{gas}$) is defined 
as f$_{\rm gas}$ = M$_{\rm gas}/$(M$_{\rm gas}$+M$_{*}$). M$_{gas}$, is computed by scaling 
the HI mass by 1.33 to account for the primordial He fraction. No correction 
is made for the molecular gas. To compute the stellar mass, $M_{*}$, the stellar 
mass to light ratio in the B band ($\Gamma_{*}$) was derived from the observed (B-V) 
colour , using from the galaxy evolution models of Bell et al.(2003) and assuming a 
``diet" Salpeter IMF. Solid triangles are from McGaugh(2005), solid squares from Verheijen(2001),
crosses from Swaters (1999) and empty triangles from C\^{o}t\'{e} et al.(2000).  Note how the GMRT 
FIGGS sample extends the coverage of all three galaxy properties.
}
\label{fig:gasfrac}
\end{figure*}

Figure~\ref{fig:hist_figgs}  shows the histogram of the absolute blue magnitude (M${\rm{_B}}$), 
distance, HI mass, and HI mass to light ratio (M${\rm{_{HI}/L_B}}$) for the FIGGS sample, while
Figure~\ref{fig:gasfrac} compares the distributions of gas fraction, luminosity and 
dynamical mass of the FIGGS galaxies with that of existing samples of galaxies with HI 
aperture synthesis observations.  The gas fraction and the dynamical masses for the FIGGS 
sample have been derived from the GMRT observations. The FIGGS 
sample has a median ${\rm{M_B}} \sim -13$ and  a median HI mass $\sim 3 \times 10^7$~M$_\odot$,
while spanning range of more than 100 in stellar light, gas mass and dynamical mass, and more 
than 4 in gas fraction. It can also be clearly seen that by focusing on fainter, lower mass
galaxies than those observed in previous HI imaging studies, FIGGS bridges the transition
to rotation dominated low mass spirals and provides a substantially extended baseline 
in mass and luminosity space for a comparative study of galaxy properties.

\begin{figure*}
\psfig{file=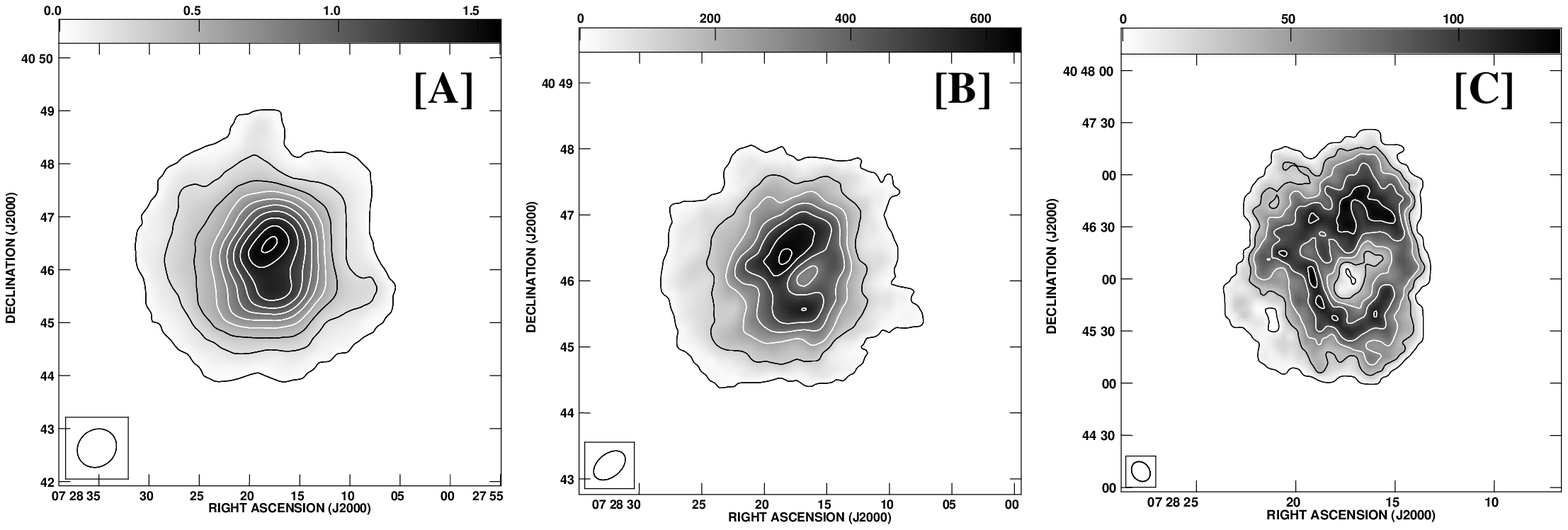,width=7.0truein}
\caption{ The figure shows the integrated HI emission from one of the galaxy in FIGGS sample, DDO 43 
at various resolutions viz. 46$^{\prime\prime} \times 42^{\prime\prime}$ (panel {\bf{[A]}}), 
32$^{\prime\prime} \times 21^{\prime\prime}$ (panel {\bf{[B]}}) and 
15$^{\prime\prime} \times 12^{\prime\prime}$ (panel {\bf{[C]}}).  The first contour level
and contour separation for these resolutions are (1.3,12.2), (2.5,18.1) and (4.0,26.2),
respectively, in units of 10$^{19}$ cm$^{-2}$.
}
\label{fig:ddo43}
\end{figure*}

\section{Science drivers for FIGGS }
\label{sec:science}

   The aim of FIGGS is to provide a large multi-wavelength database for a systematically selected 
sample of extremely faint dwarf irregular galaxies. As mentioned in Section~\ref{sec:intro}, such
a database could be used to address a diverse range of astrophysical questions. Rather than 
attempting to enumerate all of these, in this section, we describe in some detail a couple of
key science drivers for the FIGGS survey.

\subsection{Star formation and feedback in small galaxies}

     One of the main goals of FIGGS is to use the HI interferometric images in conjunction with the
optical data to study the interplay between the neutral ISM and star formation in the faintest,
lowest mass, gas rich dIrr galaxies.  The gravitational binding energy for very faint dwarf 
irregular galaxies is not much larger than the energy output from a few supernovae; consequently 
star formation in such galaxies could have a profound effect on the morphology  and  kinematics 
of the ISM of these systems.  The FIGGS data will enable us to study the ISM of most
of our sample galaxies at a linear resolution of $\sim15-300$ pc $-$ i.e. comparable to the  scales
at which energy is injected into the ISM through supernova and stellar winds. FIGGS thus  provide
a unique opportunity to study the effects of feedback from star formation in low mass, gas rich 
galaxies,  which in turn will allow us to understand the processes driving the evolution of
these galaxies. For example, it has been suggested that star formation in dwarf galaxies occurs
only above a constant threshold HI column density of N${\rm{_{HI}}}\sim10^{21}$ cm$^{-2}$
(e.g. \citealp{skillman87,taylor94}). Such a threshold could be a consequence of disk dynamics
(e.g. related to Toomre's instability criterion; \cite{rob89}) or a consequence of 
some other physical process, e.g. self shielding or thermo-gravitational instability 
(\cite{schaye04}). A preliminary study of a small subsample of FIGGS (\cite{b06}) suggested
that unlike brighter dwarfs, the faintest dwarf galaxies do not show well defined threshold
density. A detailed comparison of  H$\alpha$ and UV images with  HI column density maps for the
FIGGS sample will allow us to definitively answer the issue of the existence of a threshold
density in the faintest galaxies and also to check whether the recipes for star formation
derived from larger galaxies (\cite{rob89}) continue to be valid at this mass regime.
These are critical issues in  hierarchical galaxy formation models.

\subsection{Dark and visible matter in small galaxies}

      The second major aim of this survey is to study the relation between dark and
baryonic matter in the smallest known star forming galaxies. According to several models of 
galaxy formation and evolution, the first burst of star formation in dwarf galaxies below a 
critical halo circular velocity ($\sim$100~kms$^{-1}$) could lead to the loss of a significant 
fraction of baryons (e.g. \citealp{g00,dk03}). In fact, expulsion of gas because of energy 
input from supernovae has been postulated as a possible mechanism to produce dwarf elliptical
galaxies from gas rich progenitors (e.g. \cite{mr97}). Although a complete expulsion of 
the ISM from galaxies has not been observed so far, expansive outward motions of the 
neutral gas in dwarf galaxies has been observed in at least two galaxies (viz. 
GR8, Begum \& Chengalur 2003; NGC~625, Cannon et al. 2004). To test these models, high 
spatial resolution interferometric observations are crucial. 

   The Tully-Fisher (TF) relation demonstrates the existence of a tight relation between dark and 
luminous matter in bright spiral galaxies. \cite{mg00} (see also \cite{mg05}) showed that 
dwarf galaxies deviate from the TF relation defined by bright spirals, but that the relationship
is restored if one works with the total baryonic mass instead of the luminosity, i.e. a
``Baryonic Tully Fisher'' (BTF) relation. The FIGGS sample, both because it extends well beyond
the region of rotation dominated dwarfs and because accurate distances are known for a large
subsample, forms a very interesting dataset for studying TF and BTF relations. Most of the 
past studies have been done using the HI global velocity widths from the single dish observations 
(\citealp{geha06,mg00}). While for the brighter galaxies W$_{20}$ (the velocity width at 20\% emission,
after correction for random motions and instrumental broadening), is a good measure of the 
rotational velocity of the galaxy (Verheijen \& Sancisi 2001); it is unclear if this would
remain true in the case of faint dwarf galaxies, where random motions could be comparable 
to the peak rotational velocities (e.g. \citealp{bch03,bc04a}). For such galaxies,
it is important to accurately correct for the pressure support (``asymmetric drift''
correction) for which one needs to know both the rotation curve as well as the distribution
of the HI gas, both of which can only be obtained by interferometric observations such as
in FIGGS. The FIGGS sample would thus allow
us to concretely answer this question using actual observational data.

      The HI kinematics of FIGGS galaxies, in conjunction with the H$\alpha$ rotation curves 
can be used to accurately determine the density distribution of the dark matter halos of 
faint galaxies. Since stars generally make a minor contribution to the total mass in
the FIGGS galaxies, accurate kinematical studies can provide direct information on the 
density profiles of their dark matter halos with less uncertainties arising from the unknown 
stellar mass to light ratio. Cosmological simulations of hierarchical galaxy formation predict
a ``universal'' cusped density core for the dark matter halos of galaxies (e.g. Navarro et al. 2004).
On the other hand, observations of dIrr galaxies indicate a constant density core for their dark 
matter halos (e.g. \citealp{weldrake03,deblok03}); however this comparison remains controversial 
(e.g. \citealp{bosh01,deblok05}). FIGGS would not only provide a large sample for such a comparison,
but would also provide a data set that is less subject to uncertainties due to the unknown
stellar mass to light ratio or large scale non circular motions due to bars or spiral arms.

\section[]{HI Observations and data analysis} 
\label{sec:obs}

\begin{table*}
\begin{center}
\caption{Parameters of the GMRT observations}
\label{tab:obs}
\vskip 0.1in
\begin{tabular}{|l|cccccccc|}
\hline
Galaxy&Date of Observations&Velocity Coverage &Time on Source&synthesised Beam &  Noise &Phase Cal & Cont Noise\\
&&(\kms)&(hours)&(arcsec$^2$)&  (mJy)& &(mJy)\\
\hline
\hline
DDO 226& 8 July 2004 &257 $-$ 469& 3.5 &52$ \times46$, 26$\times 21$, 19$\times 17$& 3.2, 2.4, 2.1 &0025-260 &1.5, 0.9\\
DDO 6& 1 Feb 2004 &189 $-$ 401 & 5.0 &50$ \times 45$, 26$\times 21$, 16$\times 12$& 3.4, 1.7, 1.4 &0116-208& 1.3, 0.8 \\
UGC 685& 18 June 2004 &51 $-$ 263& 3.5 &42$\times 40$, 36$\times 25$, 27$\times 19$& 4.0, 3.5, 3.1 &0204+152& 1.9, 1.0\\
KKH 6& 9 July 2004 &$-$45 $-$ 166& 4.0 &41$\times 33$, 30$\times 20$, 16$\times 11$& 3.0, 2.6, 2.2 &0136+473 & 1.3, 0.8\\
KK 14& 19 June 2004 &317 $-$ 529& 5.0 &41$\times 38$, 28$\times 24$, 16$\times 13$& 3.0, 2.5, 2.1 &3C48 & 1.6, 0.8\\
KKH 11& 25 Nov 2004 &205 $-$ 415 & 3.0 &46$\times 36$, 21$\times 14$, 13$\times 10$& 3.0, 2.1, 1.8 &0110+565& 1.9, 0.9\\
KKH 12& 16 July 2004 &$-$ 36$-$ 176& 3.7 &47$\times 37$, 32$\times 26$, 16$\times 11$& 1.6, 1.3, 1.0 &0110+565 &1.6, 0.8\\
UGCA 92& 6 June 2005 &$-$205 $-$ 7 & 2.1 &42$\times 40$, 26$\times 21$, 16$\times 14$& 5.5, 4.2, 3.5 &0410+769 & 2.9, 1.8\\
KK 41& 8 July 2004 &$-$152 $-$ 60& 3.0 &53$\times 52$, 31$\times 22$, 20$\times 16$& 1.9, 1.3, 1.1 &0410+769& 1.3, 0.8\\
KKH 34& 9 June 2004 &$-$5 $-$ 216 & 4.5 &53$\times 50$, 34$\times 27$, 22$\times 18$& 4.0, 2.8, 2.4 &0410+769& 1.7, 0.9\\
E490-17& 17 June 2004 &404 $-$ 616 & 3.5 &49$\times 46$, 35$\times 25$, 24$\times 14$& 6.0, 5.8, 4.0 &0608-223& 2.4, 1.2\\
UGC 3755& 11 Jan 2004 &209 $-$ 421& 5.0 &42$\times 40$, 28$\times 25$, 18$\times 16$& 3.8, 3.0, 2.6 &0745+101& 1.9, 1.0\\
DDO 43& 16 Jan 2005 &248 $-$ 460& 3.5 &46$\times 42$, 32$\times 21$, 15$\times 12$& 3.2, 2.6, 2.2 &0713+438& 1.6, 1.0\\
KK 65& 25 Nov 2004 &173 $-$ 385 & 5.0 &41$\times 37$, 27$\times 25$, 19$\times 17$& 3.0, 2.3, 1.9 &0738+177& 1.7, 0.9\\
UGC 4115& 10 July 2004 &235 $-$ 447 & 4.0 &42$\times 41$, 34$\times 26$, 18$\times 14$& 3.4, 3.2, 2.7 &0745+101& 1.8, 1.1 \\
KK 69& 3 Jan 2005 &357 $-$ 569 & 4.0 &56$\times 51$, 42$\times 35$, 28$\times 24$& 4.0, 3.0, 2.5 &0741+312&1.9, 1.3\\
UGC 5186& 26 Nov 2004 &445 $-$ 657 & 5.0 &41$\times 37$, 27$\times 24$, 15$\times 13$& 3.0, 2.2, 1.6 &0958+324& 1.5, 0.8 \\
UGC 5209& 15 Jan 2005 &432 $-$ 644 & 4.5 &41$\times 37$, 27$\times 24$, 18$\times 15$& 3.2, 2.6, 2.1 &0958+324& 1.6, 0.9\\
UGC 5456& 9 July 2004 &438 $-$ 650& 3.0 &41$\times 36$, 34$\times 25$, 22$\times 20$& 4.0, 3.2, 2.6 &1008+075& 1.9, 1.2 \\
HS~117  & 8 Aug 2005 & $-$143 $-$ 69 & 3.0 & 46$\times 42$ & 3.8 & 1035+564 & 1.8, 1.1\\
UGC 6145& 11 Feb 2005 &634 $-$ 846 & 5.0 &42$\times 37$, 34$\times 27$, 20$\times 14$& 2.4, 2.0, 1.8 &1150-003& 1.4, 1.0\\
UGC 6456& 19 June 2004 &$-$208 $-$ 3& 3.5 &53$\times 39$, 23$\times 20$, 16$\times 15$& 5.0, 3.4, 3.0 &1435+760& 2.0, 1.2\\
UGC 6541& 29 Nov 2004 &144 $-$ 356 & 5.0 &42$\times 33$, 28$\times 23$, 21$\times 15$& 3.8, 3.0, 2.4 &1035+564& 2.2, 1.0 \\
KK 109& 6 June 2005 & 106 $-$ 318 & 4.0 &49$\times 41$, 27$\times 21$, 18$\times 13$& 4.8, 3.5, 2.7 &1227+365& 2.3, 1.2 \\
DDO 99& 30 June 2005 &136 $-$ 348& 3.5 &45$\times 37$, 28$\times 23$, 19$\times 16$& 3.8, 3.0, 2.7 &1227+365& 2.0, 1.1 \\
E379-07& 19 Jan 2005 &534 $-$ 746 & 4.5 &54$\times 47$, 33$\times 24$, 19$\times 18$& 3.0, 2.0, 1.7 &1154-350& 1.8, 1.1\\
KK 127 & 9 July 2004 & 46.0 $-$ 258 & 3.5 & 49$\times 44$ & 3.0 &  1227+365 & 1.7, 1.1\\ 
E321-014& 7 Oct 2005 &507 $-$ 719 & 3.5 &51$\times 45$, 28$\times 18$, 15$\times 10$& 3.9, 2.7, 2.3 &1154-350& 1.8, 0.9 \\
UGC 7242& 3 Feb 2004 &$-$37 $-$ 174 & 6.5 &45$\times 37$, 27$\times 23$, 18$\times 15$& 2.2, 1.9, 1.6 &1313+675& 1.4, 0.8 \\
UGC 7505& 28 Nov 2004 &210 $-$ 422& 3.0 &45$\times 39$, 28$\times 27$, 24$\times 18$& 4.0, 3.2, 2.8 &1227+365& 1.6, 1.1 \\
KK 144& 12 July 2004 &377 $-$ 589& 4.5 &41$\times 40$, 28$\times 24$, 19$\times 14$& 4.0, 3.2, 2.7 &1221+282& 1.8, 1.0 \\
DDO 125& 6 June 2005 &89 $-$ 301 & 4.0 &45$\times 36$, 31$\times 22$, 20$\times 14$& 4.2, 3.8, 3.0 &1227+365& 1.8, 1.2 \\
UGC 7605& 1 Feb 2004 &204 $-$ 416 & 7.0 &43$\times 38$, 29$\times 24$, 16$\times 12$& 2.3, 2.0, 1.7 &1227+365& 1.5, 0.9 \\
UGC 8055& 13 June 2005 &512 $-$ 724 & 6.0 &40$\times 37$, 27$\times 25$, 19$\times 17$& 3.3, 2.7, 2.4 &1254+116& 2.0, 1.2\\
UGC 8215& 29 Nov 2004 &112 $-$ 324 & 6.0 &46$\times 39$, 28$\times 22$, 17$\times 14$& 3.6, 2.7, 2.4 & 1227+365& 2.1, 1.2\\
DDO 167& 10 July 2004 & 57 $-$ 269& 3.5 & 51$\times 38$, 29$\times 23$, 19$\times 16$&  6.0, 4.7, 4.1 & 1227+365& 2.5, 1.3\\
KK 195& 4 Jan 2005 &460 $-$ 672 & 4.5 & 62$\times 54$, 30$\times 24$, 18$\times 16$&  3.9, 2.7, 2.1 & 1018-317&2.2, 1.1 \\
KK 200& 26 Nov 2004 &381 $-$ 593 & 5.0 & 48$\times 47$, 32$\times 23$, 21$\times 17$&  2.9, 2.1, 1.8 &1316-336& 1.8, 0.9\\
UGC 8508& 31 Jan 2004 &$-$44 $-$ 167& 7.0 & 42$\times 38$, 32$\times 24$, 18$\times 15$&  2.6, 2.2, 1.8 &1400+621& 1.5, 0.8\\
E444-78& 20 June 2004 &475 $-$ 686 & 2.5 & 48$\times 47$, 26$\times 21$, 18$\times 11$&  6.0, 4.0, 3.5 &1316-336& 2.3, 1.6\\
UGC 8638& 9 July 2004 &168 $-$ 380 & 2.5 & 44$\times 36$, 25$\times 17$, 16$\times 11$&   4.0, 2.6, 2.0 &1330+251& 1.7, 1.0 \\
DDO 181& 6 June 2005 &96 $-$ 308& 5.5 & 50$\times 41$, 26$\times 21$, 17$\times 14$&   5.2, 3.4, 2.7 &3C286& 2.1, 1.2 \\
I4316& 7 Aug 2005 &474 $-$ 686& 2.7 & 48$\times 46$, 26$\times 20$, 15$\times 11$&   3.6, 2.8, 2.3 &1316-336 & 1.9, 1.0\\
DDO 183& 31 Jan 2004 &86 $-$ 298& 6.5 & 42$\times 38$, 31$\times 24$, 17$\times 13$&   2.7, 2.0, 1.7 &1331+305& 1.5, 0.9\\
UGC 8833& 16 June 2004 &121 $-$ 333 & 3.5 & 41$\times 39$, 30$\times 25$, 21$\times 18$&   3.7, 2.8, 2.3 &3C286& 2.0, 1.5 \\
DDO 187& 16 June 2004 &47 $-$ 259 & 2.5 & 46$\times 37$, 30$\times 25$, 19$\times 13$&   5.0, 4.1, 3.4 &3C286& 2.4, 1.5\\
P51659& 14 Jan 2005 &285 $-$ 497 & 3.0 & 48$\times 41$, 26$\times 20$, 15$\times 14$&   3.4, 2.6, 2.2 &1316-336& 1.9, 1.1\\
KK 246& 16 June 2004 &255 $-$ 469 & 2.5 & 61$\times 39$, 36$\times 21$, 15$\times 11$&   4.2, 3.5, 2.9 &1923-210& 2.2, 1.2\\
%UGCA 438& 12 June 2004 &$-$43 $-$ 168& 2.0 & &    &2302-373\\
KKH 98& 10 July 2004 &$-$243 $-$ $-$32 & 6.0 & 42$\times 41$, 32$\times 28$, 15$\times 14$&   3.2, 2.7, 2.4 &0029+349& 2.1, 1.2\\
\hline
\end{tabular}
\end{center}
\end{table*}

For all the GMRT HI observations, the observing bandwidth of 1~MHz was divided into  
128~spectral channels, yielding a spectral resolution of 7.81~kHz (velocity resolution 
of 1.65~\kms).  It is worth noting that this velocity resolution is $\sim$ 4 times better
than most earlier  interferometric studies of such faint dwarf galaxies (e.g.\cite{lo93}). 
This high velocity  resolution is crucial to detect large scale velocity gradients in the
faintest dwarf galaxies (e.g. \cite{bch03,bc04a}). For each observing run, absolute flux 
and bandpass calibration was done by observing one of the standard flux calibrators 3C48,
3C286 and 3C147, at the start and end of the observations. For the sample galaxies with 
low LSR velocities, particular care was taken to choose a  bandpass calibrator which 
does not have any absorption feature in the relevant velocity range. The phase 
calibration was done once every 30 min by observing a nearby VLA phase calibrator source. 

\begin{figure}
\psfig{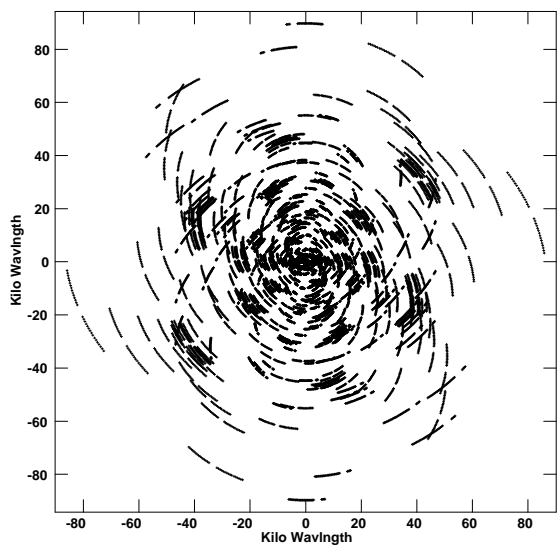}
\caption{ The figure  shows the (u,v) coverage for 
UA~92, the sample galaxy with the shortest on source integration time (viz. 2.1 hours)
}
\label{fig:uvcov}
\end{figure}

     The GMRT data were reduced in the usual way using the standard tasks in classic AIPS.
For each run, bad visibility points were edited out, after which the data were
calibrated. The GMRT does not do online doppler tracking --  any required doppler
shifts have to be applied during the offline analysis. However, for all of the sample
galaxies, the differential doppler shift over our observing interval was much less than
the channel width, hence, there was no need to apply any offline correction. 
The GMRT has a hybrid configuration (Swarup et al. 1991) with 14 of its
30 antennas located in a central compact array with size $\approx$ 1 km
($\approx$ 5 k$\lambda$ at 21cm) and  the remaining antennas distributed
in a roughly ``Y'' shaped configuration, giving a maximum baseline length
of $\approx$ 25 km ($\approx$ 120 k$\lambda$ at 21 cm). The baselines
obtained from antennas in the central compact array are similar in length to
those of the ``D'' array of the VLA, while the baselines between the arm
antennas are comparable in length to the ``B'' array of the VLA. A single
observation with the GMRT hence yields information on both large and small
angular scales. 
Data cubes at a range of angular resolutions were made using appropriate 
(u,v) ranges and tapers. In this paper we present only the low resolution 
HI images, i.e. made using (u,v) ranges of $\sim$~0$-$5~k$\lambda$, 0$-$10~k$\lambda$ 
and 0$-20$~k$\lambda$. Higher resolution observations of the FIGGS sample will be 
presented in the companion paper.  
To obtain the low resolution HI images for the sample galaxies, the uv-taper 
at each (u,v) range was adjusted to achieve as close as possible to a circular 
synthesized beam. A low resolution data cube was generated for each galaxy, using 
the AIPS task IMAGR, and the individual channels were inspected using the task
TVMOVIE to identify the channels with HI emission. Emission was detected from all
of the galaxies in our sample, except for SC~24, HS~117, KK~127 and KKR~25. Apart from
HS 117, all of these galaxies were previously claimed to be 
detected by single dish observations. The GMRT data suggest that the previous 
flux measurements were spurious, probably as a result of confusion with galactic
emission. The galaxies KK~127 and SC~24 are likely to be  distant  dwarf irregular 
galaxies  whereas KKR~25 is a normal dwarf spheroidal galaxy (\cite{bc05,kk06}).
In the case of HS~117, single dish observations did not detect this galaxy (\cite{hs98}).
The HI data given in Karachentsev et al. (2002) is a result of
misidentifying galactic HI emission as emission from HS~117.
For the rest of the galaxies in the sample,  frequency channels with emission 
were identified and the continuum maps were made at  both low (26$^{\prime\prime} 
\times 22^{\prime\prime}$) and high (5$^{\prime\prime} \times 5^{\prime\prime}$) 
resolutions  using the average of the remaining line free channels. No extended  
or compact  emission was detected from the disk of any of our sample  galaxies. 
All other continuum sources lying with the field of view were  subtracted using 
the task UVSUB. After continuum subtraction, deconvolved data cubes of the line
emission were made at a range of resolutions using the AIPS task IMAGR.

 HI images at both high and low spatial resolutions are crucial for a complete 
understanding of the properties of the atomic ISM of faint dwarf galaxies. As an 
example, Figure~\ref{fig:ddo43} shows the integrated HI column distribution at 
various resolutions for one of the FIGGS galaxies  DDO 43. This galaxy shows a faint, 
extended HI envelope  which is only seen clearly in the lowest resolution HI maps. 
On the other hand, DDO 43 also has a large hole in the center (see also the
VLA observations in \cite{simpson05}), which is seen  in the high resolution 
HI map. However this hole in the HI distribution is not at all obvious in the
low resolution HI maps due to the beam smearing.

The setup and observational results for 49 galaxies from the FIGGS sample are given 
in Table~\ref{tab:obs}.  For the remaining 15 sample galaxies,  the details of the 
observations and data analysis can be found in \citealp{bch03,bc03, bc04a,bc04b,
bcks05,bc05,bck05,b06} and Chengalur et al. 2008 (in preparation). 
In the case of UGCA~438, most of  the short baselines were missing because of the non availability
of some of the GMRT antennas during the observing run, thus missing the 
diffuse, extended emission from the galaxy. Future observations of this 
galaxy are planned. We have not considered this galaxy for the 
analysis  in this paper. 
The columns in Table~\ref{tab:obs} are as follows: Column(1)~the galaxy name, 
Column(2)~the date of observations, Column(3)~the velocity coverage of the observation, 
Column(4)~the total integration time on source, Column(5)~the synthesized beam sizes
of the data cubes, Column(6)~the rms noise per channel for the different resolution
data cubes, Column(7)~the phase calibrator used, Column(8)~the $3\sigma$ limits on
continuum emission from the galaxy at  resolutions of (26$^{\prime\prime} \times 
22^{\prime\prime}$) and (5$^{\prime\prime} \times 5^{\prime\prime}$) respectively.
We note that although for some of the sample galaxies the on-source integration time
is short ($\sim 2 - 2.5$~hours), the hybrid configuration of the GMRT leads to a reasonable
sampling of the (u,v) plane. As  an example, Figure~\ref{fig:uvcov}  shows the (u,v) coverage
for UA~92, the sample galaxy with the shortest on source integration time (viz. 2.1 hours).

        We examined the line profiles at various locations in the galaxy
and found that they were (to zeroth order) symmetric and  single peaked.
For some galaxies, in the very high column density regions, a double gaussian and/or
a gauss-hermite fit does provide a somewhat better description of the data,
but even in these regions, the mean velocity produced by the moment
method agrees within the errors with the peak velocity of the profile.
Since we are interested here mainly in the systematic velocities, moment
maps provide an adequate description of the data.  Moment maps (i.e.
maps of the total integrated flux (moment~0),  the flux weighted velocity
(moment~1) and the flux weighted velocity dispersion (moment~2)) were
made from the data cubes using  the AIPS task MOMNT. To obtain the moment
maps, lines of sight with a low signal to noise ratio were excluded by
applying a cutoff at the $2\sigma$ level, ($\sigma$ being the rms noise level
in a line free channel), after smoothing in velocity (using boxcar
smoothing three channels wide) and position (using a gaussian with
full width at half maximum (FWHM)~$\sim 2$ times that of the synthesized beam). 
Maps of the velocity field were also made in GIPSY using single
gaussian fits to the individual profiles. The velocities produced by
MOMNT in AIPS are in reasonable agreement with those obtained using a single
gaussian fit. Note that the AIPS moment~2 map systematically underestimates
the velocity dispersion (as obtained from gaussian fitting) particularly
near the edges where the signal to noise ratio is low. This can be
understood as the effect of the thresholding algorithm used by the MOMNT
task to identify the regions with signal.

\begin{figure*}
\psfig{file=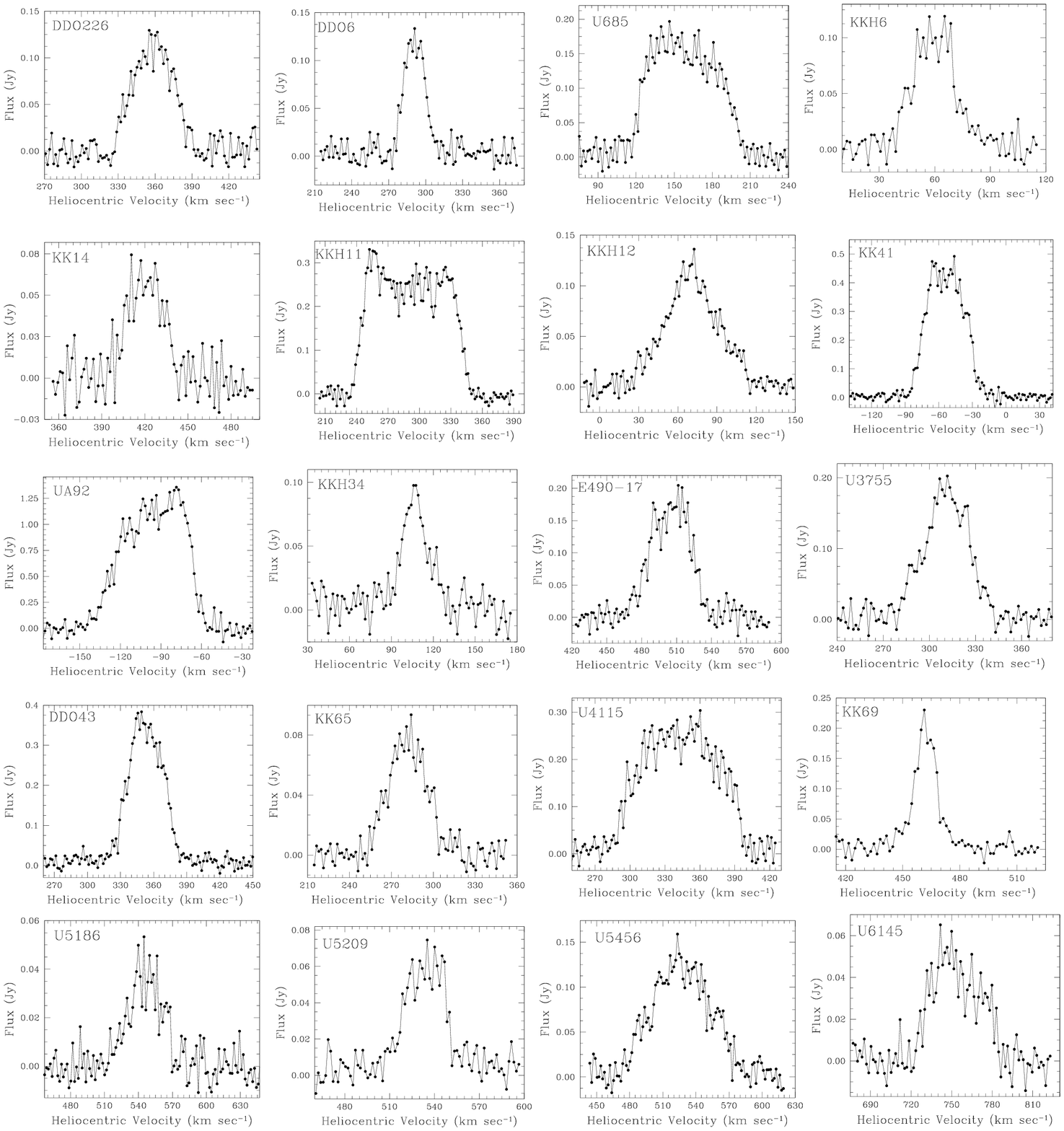,width=7.0truein}
\caption{The global HI profiles of the sample galaxies
obtained from the lowest resolution data  cubes (see Table~\ref{tab:obs}).
}
\label{fig:spec}
\end{figure*}

\begin{figure*}
\addtocounter{figure}{-1}
\psfig{file=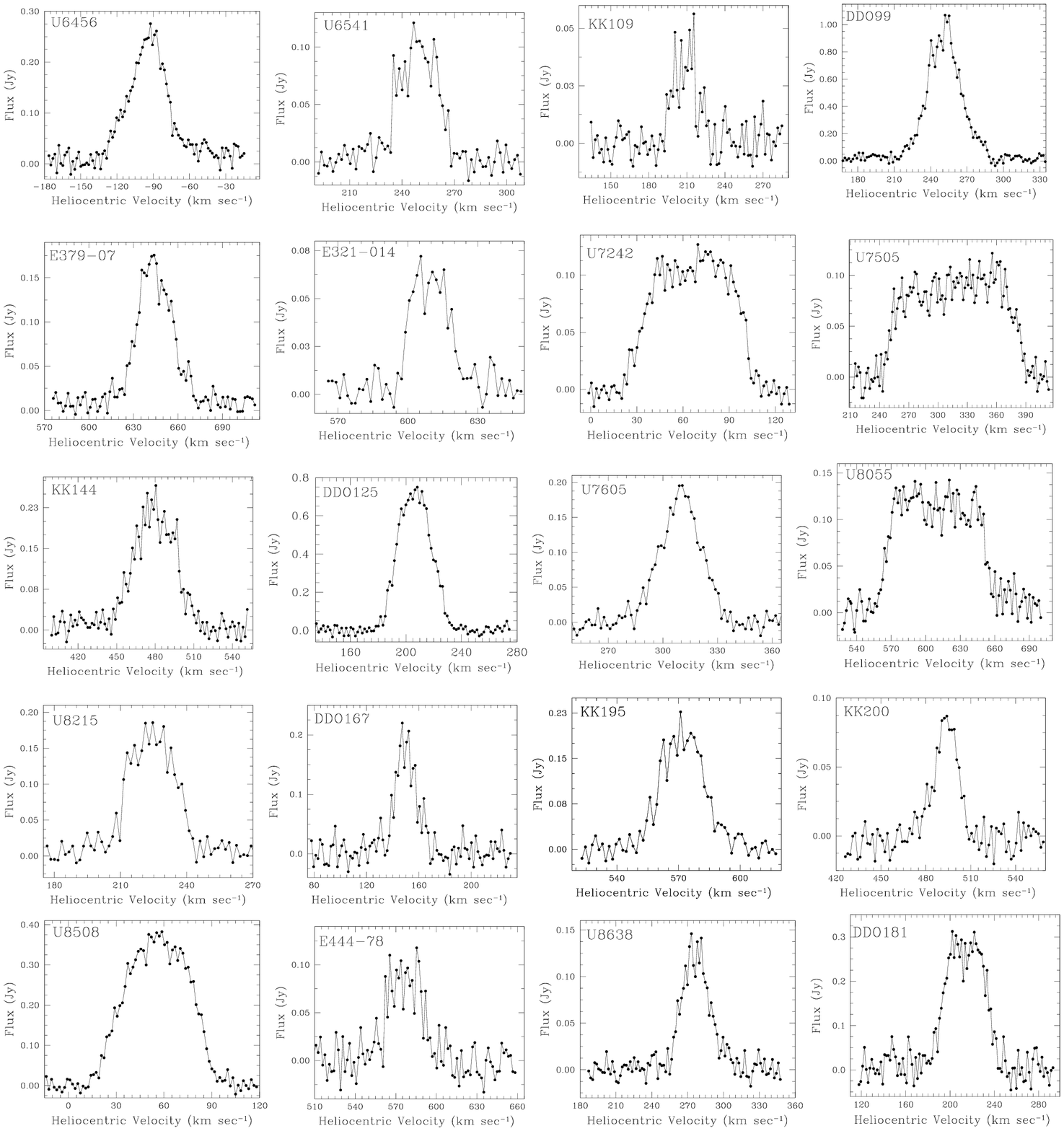,width=7.0truein}
\caption[Continued]{({\it{continued}})The global HI profiles of the sample galaxies
obtained from the lowest resolution data  cubes (see Table~\ref{tab:obs}).
}
%\label{fig:vrot}
\end{figure*}

\begin{figure*}
\addtocounter{figure}{-1}
\psfig{file=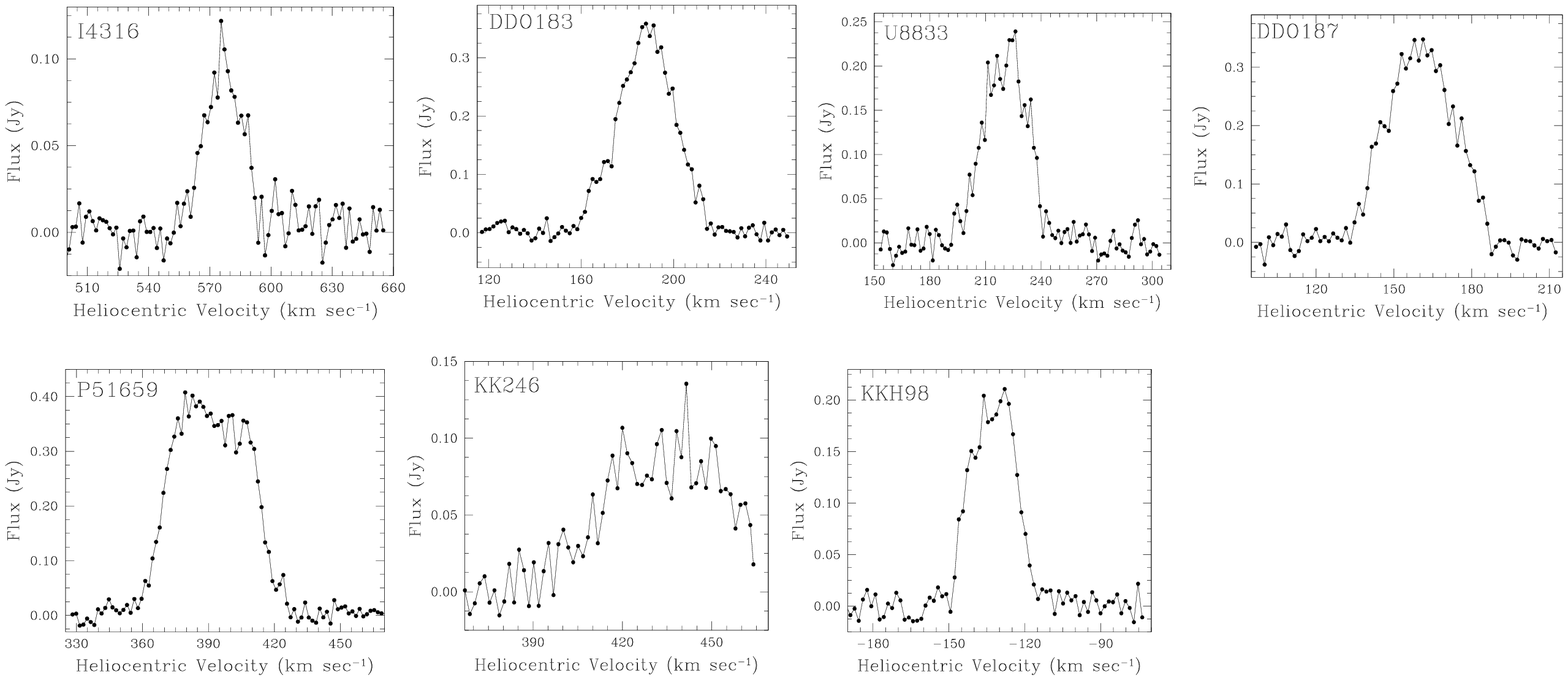,width=7.0truein}
\caption[Continued]{({\it{continued}})The global HI profiles of the sample galaxies
obtained from the lowest resolution data  cubes (see Table~\ref{tab:obs}).
}
%\label{fig:vrot}
\end{figure*}

\begin{figure*}
\psfig{file=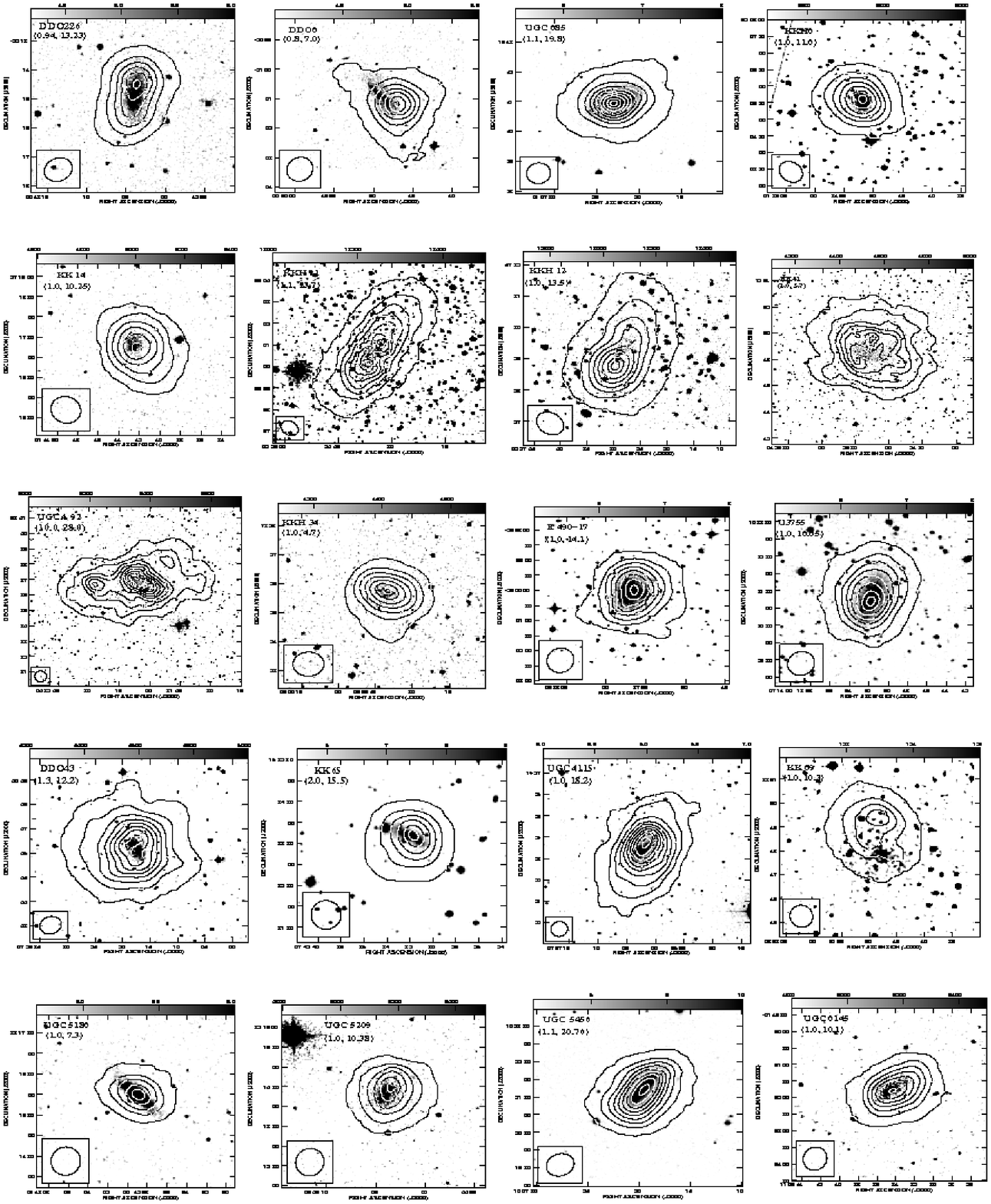,width=7.0truein}
\caption{The GMRT integrated HI column density distribution (contours) overlayed on the
optical DSS images (grey scales) of FIGGS galaxies  from the lowest resolution data  cubes (see Table~\ref{tab:obs}).
The contours are uniformly spaced. The first contour level and the contour separation are printed below the 
galaxy name in units of 10$^{19}$ cm$^{-2}$.
}
\label{fig:mom0}
\end{figure*}

\begin{figure*}
\addtocounter{figure}{-1}
\psfig{file=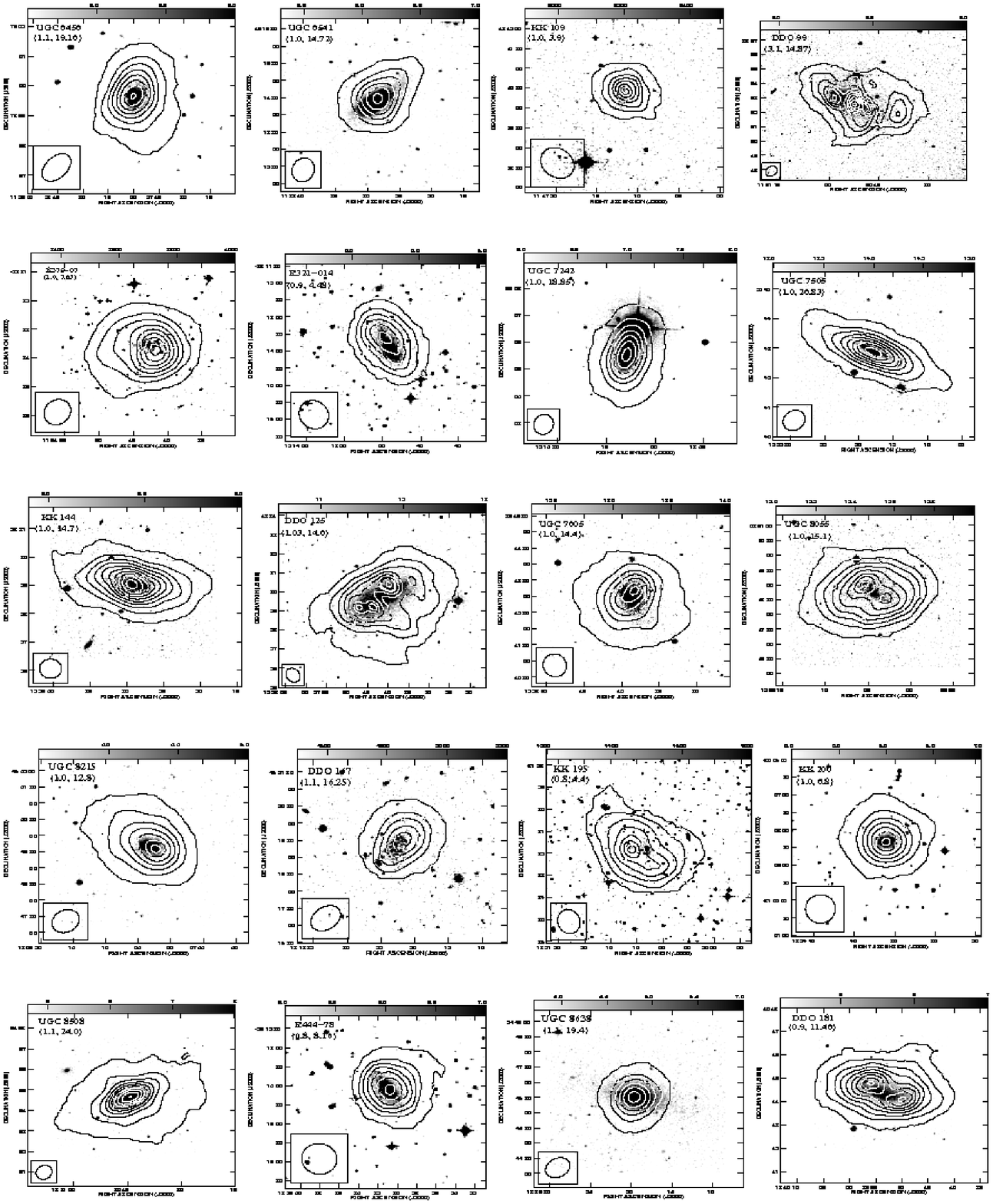,width=7.0truein}
\caption[Continued]{({\it{continued}}) The GMRT integrated HI column density distribution
(contours) overlayed on the optical DSS images (grey scales) 
of FIGGS galaxies from the lowest resolution data
cubes (see Table~\ref{tab:obs}).The first contour level and the contour separation are printed below the
galaxy name in units of 10$^{19}$ cm$^{-2}$.
}
%\label{fig:vrot}
\end{figure*}

\begin{figure*}
\addtocounter{figure}{-1}
\psfig{file=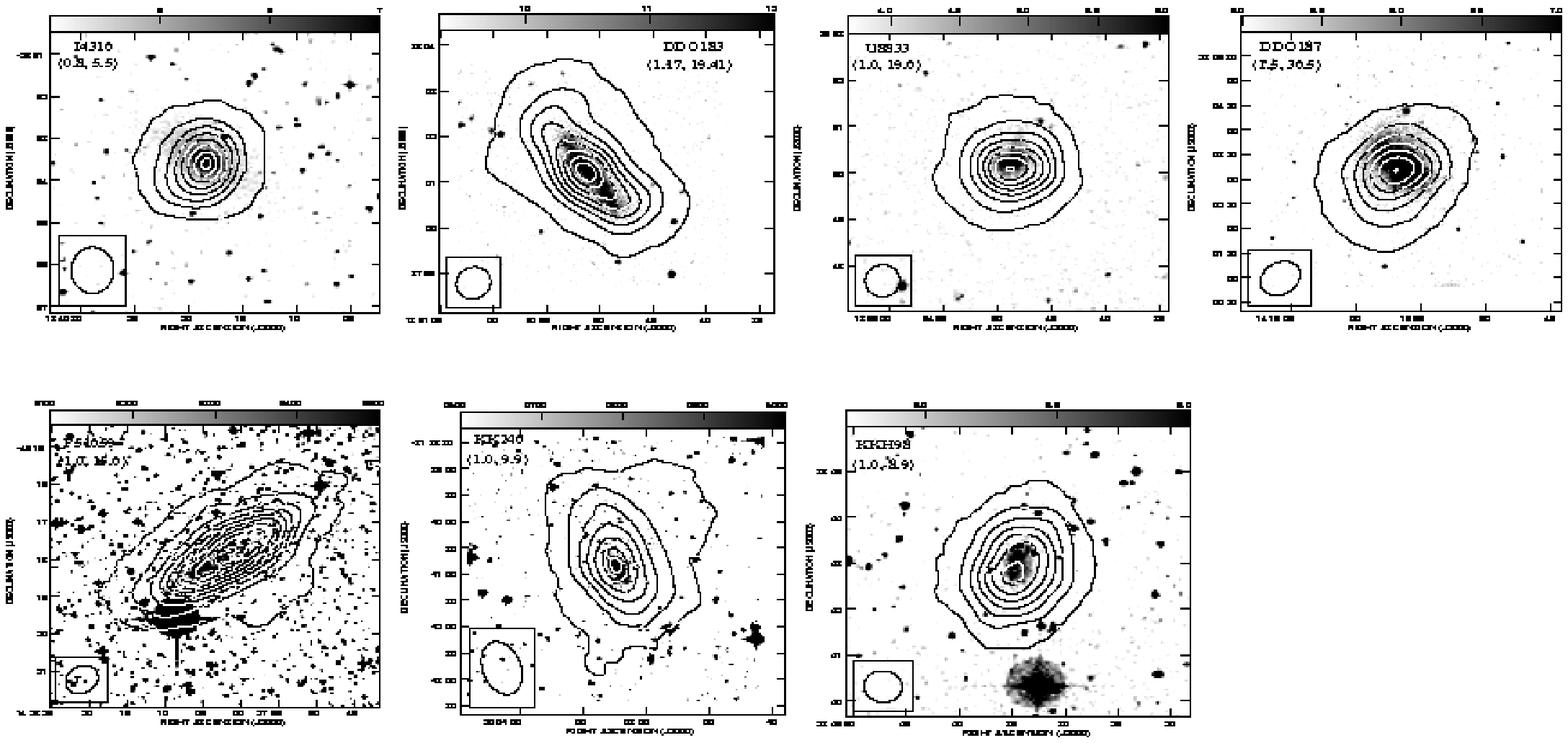,width=7.0truein}
\caption[Continued]{({\it{continued}}) The GMRT integrated HI column density distribution
(contours) overlayed on the optical DSS images (grey scales) 
of FIGGS galaxies from the lowest resolution data
cubes (see Table~\ref{tab:obs}).
The first contour level and the contour separation are printed below the
galaxy name in units of 10$^{19}$ cm$^{-2}$.
}
%\label{fig:vrot}
\end{figure*}

\section[]{Results and Discussion}
\label{sec:result}

A detailed analysis of FIGGS data will be presented in companion papers. Here we restrict ourselves
to a preliminary analysis of the global HI and optical properties of the FIGGS sample.

\begin{table*}
\caption{Results from the GMRT observations}
\label{tab:hiresult}
\begin{tabular}{|l|c|c|c|c|c|c|c|c|c|c|}
\hline
  \multicolumn{1}{c|}{Galaxy} &
  \multicolumn{1}{|c|}{FI${\rm{_{GMRT}}}$} &
  \multicolumn{1}{c|}{V${\rm{_{sys}}}$} &
  \multicolumn{1}{c|}{$\Delta$V$_{50}$} &
  \multicolumn{1}{c|}{D${\rm{_{HI}}}$} &
  \multicolumn{1}{c|}{M${\rm{_{HI}}}$} &
  \multicolumn{1}{c|}{${\rm{\frac{M_{HI}}{L_B}}}$} &
  \multicolumn{1}{c|}{$\frac{D_{HI}}{D_{Ho}}$} &
  \multicolumn{1}{c|}{${\rm{\frac{FI_{GMRT}}{FI_{SD}}}}$}& 
  \multicolumn{1}{c|}{i${\rm{_{HI}}}$} & 
  \multicolumn{1}{c|}{Ref} \\
 & (Jy kms$^{-1}$)& (kms$^{-1}$)&(kms$^{-1}$)&($\prime$)&(10$^{6} M_\odot$)& & & &(deg)&\\
\hline
\hline
  And IV &   19.5 $\pm$ 2.0  &                237.0 & 90.0     &    7.6   & 205.19   & 16.9  & 6.9     &  0.87 $\pm$ 0.11     &      55.0 $\pm$ 5.0 &14  \\ 
  DDO 226      & 4.8 $\pm$ 0.5   & 358.57 & 37.0 & 3.5 & 25.95 & 0.36 & 1.09  & 0.79 $\pm$ 0.10& 55.0 $\pm$ 5.0&1\\
  DDO 6        & 2.6 $\pm$  0.3  & 291.83 & 19.2 & 3.3 & 6.82 & 0.44 & 1.57   & 0.77 $\pm$ 0.10& $-$&2 \\
  UGC 685        & 11.8 $\pm$ 1.2  & 156.29 & 64.4 & 3.6 & 56.15 & 0.68 & 1.64 & 0.99 $\pm$ 0.11& 36.0 $\pm$ 4.0&3\\
  KKH 6        & 3.0 $\pm$ 0.3   & 59.92 & 28.0 & 2.6 & 10.18 & 0.71 & 2.90   & 0.72 & 30.0 $\pm$ 3.0& 2\\
  KK 14        & 1.8 $\pm$ 0.2   & 420.11 & 27.7 & 2.4 & 21.93 & 1.98 & 1.50  & 0.89 $\pm$ 0.10 & 45.0 $\pm$ 3.0& 6\\
  KKH 11       & 25.0 $\pm$ 2.5  & 295.71 & 84.4 & 7.2 & 52.88 & 1.56 & 4.23 & 1.09 $\pm$ 0.12 & 66.0 $\pm$ 3.0& 2\\
  KKH 12       & 5.5 $\pm$ 0.6   & 70.0 & 48.4 & 4.6 & 11.63 & 0.46 & 2.10    & 0.34& 60.0 $\pm$ 3.0& 2\\
  KK 41        & 18.8 $\pm$ 1.9  & $-$54.2 & 38.5 & 8.7 & 67.20 & 1.03 & 3.35 & 0.42 & 30.0 $\pm$ 4.0&2 \\
  UGCA 92        & 70.0 $\pm$ 7.1  & $-9$4.58 & 56.2 & 9.0 & 156.05 & 0.55 & 4.50 & 0.70& 56.0 $\pm$ 3.0 & 5\\
  KK 44        & 4.6 $\pm$ 0.4   & 77.5 & 21.4& 3.2& 12.2 & 1.4& 2.3& 1.02 $\pm$ 0.11& 61.0 $\pm$ 5.0 & 11\\
  KKH 34       & 2.1 $\pm$ 0.2   & 106.29 & 21.7 & 2.6 & 10.44 & 0.81 & 2.60 & 0.89 $\pm$ 0.1& 45.0 $\pm$ 3.0&2 \\
  E490-17     & 7.3 $\pm$ 0.7   & 505.17 & 39.2 & 3.0 & 30.26 & 0.32 & 1.50 & 1.11 $\pm$ 0.13 & $-$&2 \\
  UGC 3755       & 6.5 $\pm$ 0.7   & 310.81 & 34.5 & 3.0 & 41.30 & 0.29 & 1.67 & 0.96 $\pm$ 0.11 & 46.0 $\pm$ 4.0& 2\\
  DDO 43       & 14.2  $\pm$ 1.4 & 352.63 & 36.5 & 5.0 & 203.02 & 1.64 & 2.78 & 1.20 $\pm$ 0.15 & 30.0 $\pm$ 5.0&2\\
  KK 65        & 2.52 $\pm$ 0.3  & 281.45 & 33.3 & 2.1 & 34.38 & 0.43 & 2.33 & 0.97 $\pm$ 0.11& 47.0 $\pm$ 4.0&2 \\
  UGC 4115       & 21.6 $\pm$ 2.2  & 342.78 & 78.0 & 6.0 & 285.53 & 3.59 & 4.00 & 1.00 $\pm$ 0.11& 50.0 $\pm$ 7.0& 2\\
  KDG 52       & 3.8 $\pm$ 0.4   & 116.0 & 20.6 & 3.5 & 10.8 & 1.8 & 2.7 & 0.85 $\pm$ 0.11 & 23.0 $\pm$ 4.0 & 11 \\
 UGC 4459        & 21.5 $\pm$ 2.2  & 19.2 & 29.6 & 4.5 & 64.2 & 1.4 & 2.8 &1.01 $\pm$ 0.11 & 30.0 $\pm$ 4.0 & 11\\
  KK 69        & 3.0 $\pm$ 0.3   & 462.04 & 13.1 & 4.0 & 41.89 & 2.11 & 2.0 & 1.07 $\pm$ 0.12 & 35.0 $\pm$ 3.0&2 \\
  UGC 5186       & 1.4 $\pm$ 0.1   & 546.08 & 34.0 & 1.6 & 15.66 & 0.65 & 1.00 & 0.96 $\pm$ 0.11 &$-$& 2\\
  UGC 5209       & 2.0 $\pm$ 0.2   & 535.19 & 31.6 & 1.9 & 21.10 & 0.75 & 2.11 & 1.21 $\pm$ 0.13 & $-$& 7\\
  UGC 5456       & 8.0 $\pm$ 0.8 & 526.75 & 62.4 & 2.8 & 58.96 & 0.35 & 1.50 & 1.16 $\pm$ 0.14& 60.0 $\pm$ 5.0& 3\\
  UGC 6145       & 2.1 $\pm$ 0.2 & 753.0 & 41.1 & 2.7 & 27.02 & 0.96 & 1.59 & 1.00$ \pm$ 0.11& 55.0 $\pm$ 5.0& 5\\
  UGC 6456       & 10.1 $\pm$ 1.0 & $-$93.69 & 37.4 & 3.7 & 43.89 & 0.69 & 2.47 & 0.72& 65.0 $\pm$ 3.0&2\\
  UGC 6541       & 2.7 $\pm$ 0.3 & 249.36 & 25.5 & 2.1 & 9.65 & 0.20 & 1.29 & 0.90 $\pm$ 0.10 & $-$&2\\
  NGC 3741       & 74.7 $\pm$ 7.5 & 228.8 & 83.4 & 14.6 & 130.0 & 4.7 & 8.80 & $-$ &68.0 $\pm$ 4.0&  10\\
  KK 109       & 0.76 $\pm$ 0.08 & 210.67 & 18.2 & 1.4 & 3.62 & 2.98 & 1.00 & 1.08 $\pm$ 0.12 &$-$&2 \\
  DDO 99       & 33.0 $\pm$ 3.3 & 251.22 & 33.7 & 9.6 & 52.42 & 1.32 & 2.74 & 1.04 $\pm$ 0.12& $-$&5\\
  E379-07     & 5.0 $\pm$ 0.5 & 644.04 & 28.5 & 3.6 & 31.77 & 2.43 & 3.27 & 0.93 $\pm$ 0.10& 31.0 $-$ 6.0& 8\\
  E321-014    & 1.3 $\pm$ 0.1 & 609.39 & 19.0 & 2.1 & 3.13 & 0.17 & 1.49 & 0.46 $\pm$ 0.05& $-$& 5\\
  UGC 7242       & 7.2 $\pm$ 0.7 & 66.05 & 66.5 & 4.0 & 45.75 & 0.70 & 2.11 & 1.02 $\pm$ 0.11& 58.0 $\pm$ 3.0& 5\\
 CGCG 269-049 & 4.7 $\pm$ 0.5 & 159.0 & 26.6 & 2.6 & 26.4 & 0.9 & 2.3 & 0.91 $\pm$ 0.10 & 42.0 $\pm$ 4.0 & 11,12\\
 UGC 7298        & 5.2 $\pm$ 0.5 & 174.0 & 21.4 & 3.5 & 21.6 & 1.7 & 3.1 & 1.06 $\pm$ 0.11 & 28.0 $\pm$ 3.0 & 11 \\
  UGC 7505       & 11.5 $\pm$ 1.2 & 316.0 & 125.1 & 5.3 & 442.78 & 1.72 & 5.30 & 0.92 $\pm$ 0.10& 70.0 $\pm$ 4.0&5 \\
  KK 144       & 8.7 $\pm$ 0.9 & 479.54 & 37.5 & 4.7 & 81.15 & 4.80 & 3.13 & 1.01 $\pm$ 0.11 & 57.0 $\pm$ 4.0& 2\\
  DDO 125      & 21.7 $\pm$ 2.2 & 206.25 & 27.4 & 7.0 & 31.87 & 0.44 & 1.67 & 1.00 $\pm$ 0.11& $-$&5\\
  UGC 7605       & 4.93 $\pm$ 0.5 & 309.95 & 25.8 & 3.3 & 22.29 & 0.55 & 1.50 & 0.87 $\pm$ 0.10 & 40.0 $\pm$ 5.0&2 \\
  UGC 8055       & 11.0 $\pm$ 1.1 & 609.05 & 85.6 & 4.2 & 782.64 & 3.20 & 3.00 & 1.34 $\pm$ 0.32& 45.0 $\pm$ 3.0&3 \\
  GR 8         & 9.0 $\pm$ 0.9 & 217.0 & 26.0 & 4.3 & 10.38 & 1.02 & 2.3 & 1.03 $\pm$ 0.11 & 27.0 $\pm$ 4.0& 11\\
  UGC 8215       & 4.5 $\pm$ 0.5 & 224.15 & 24.6 & 3.5 & 21.41 & 1.72 & 3.50 & 1.05 $\pm$ 0.11& 45.0 $\pm$ 4.0&2 \\
  DDO 167      & 3.7 $\pm$ 0.4 & 150.24 & 18.6 & 2.0 & 14.51 & 0.78 & 1.25 & 0.88 $\pm$ 0.10 & $-$&4 \\
  KK 195       & 4.8 $\pm$ 0.5 & 571.91 & 24.0 & 5.0 & 30.50 & 3.88 & 3.85 & 0.91 $\pm$ 0.11& 52.0 $-$ 4.0& 5\\
  KK 200       & 1.6 $\pm$ 0.2 & 493.69 & 17.4 & 1.4 & 7.96 & 0.84 & 1.00 & 0.94 $\pm$ 0.11 &$-$& 2\\
  UGC 8508       & 18.3 $\pm$ 1.8 & 56.17 & 45.8 & 6.6 & 29.07 & 1.21 & 3.30 & 1.21 $\pm$ 0.14& 53.0 $\pm$ 4.0&4 \\
  E444-78     & 2.3 $\pm$ 0.2 & 577.0 & 30.6 & 0.9 & 14.62 & 0.45 & 1.67 & 0.83 $\pm$ 0.12& 42.0 $-$ 3.0& 9\\
  UGC 8638       & 3.5 $\pm$ 0.4 & 275.9 & 30.8 & 1.2 & 13.76 & 0.30 & 1.00 & 0.90 $\pm$ 0.10 & $-$&2 \\
  DDO 181      & 12.2 $\pm$ 1.2 & 213.6 & 39.1 & 5.2 & 27.55 & 1.08 & 3.25 & 1.07 $\pm$ 0.12&53.0 $\pm$ 3.0&2\\
  I4316       & 2.2 $\pm$ 0.2 & 576.34 & 21.5 & 2.8 & 10.01 & 0.18 & 1.00 & 1.05 $\pm$ 0.12 & $-$ &5\\
  DDO 183      & 10.5 $\pm$ 1.1 & 188.37 & 28.7 & 4.6 & 25.90 & 0.90 & 2.71 & 1.07 $\pm$ 0.12& 67.0 $\pm$ 3.0 &2\\
  UGC 8833       & 6.3 $\pm$ 0.6 & 221.03 & 27.8 & 3.0 & 15.16 & 1.05 & 2.31 & 1.05 $\pm$ 0.11& 26.0 $\pm$ 3.0&2 \\
  KK 230       & 2.2 $\pm$ 0.2 & 63.31 & 17.0 & 3.0 & 1.90 & 1.9 & 3.3 & 0.86 $\pm$ 0.11 & 50.0 $\pm$ 4.0 & 11\\ 
  DDO 187      & 11.1 $\pm$ 1.1 & 159.95 & 30.6 & 3.4 & 16.30 & 1.04 & 1.36 & 0.93 $\pm$0.10&37.0 $\pm$ 4.0&2\\
  P51659      & 17.4 $\pm$ 1.7 & 391.48 & 46.4 & 6.5 & 52.99 & 6.31 & 2.71 & 1.03 $\pm$ 0.11& 68.0 $\pm$ 4.0&5 \\
  KK 246       & 4.4 $\pm$ 0.4 & 434.71 & 52.2 & 3.5 & 84.50 & 1.36 & 2.69 & 0.53 & 56.0 $\pm$ 3.0&5 \\
  %UA438      & 2.4 $\pm$ 0.2 & 59.26 & 8.07 & 4.0 & 2.73 & 0.12 & 1.67 & 0.16\\
  KK 250       & 16.4 $\pm$ 1.6 & 126.0 & 95.5 & 5.8 & 121.0 & 1.2 & 3.2 & 0.82 $\pm$ 0.11 & 73.0 $\pm$ 4.0 & 13\\
  KK 251       & 10.6 $\pm$ 1.0 & 130.3 &  51.7 & 4.2 & 78.0 & 1.6 & 2.6 & 0.73 $\pm$0.11 &  59.0 $\pm$ 5.0 & 13 \\
 DDO 210       & 12.1 $\pm$ 1.2 & $-$139.5 & 19.1 & 4.8 & 2.8 & 1.0 & 1.3 & 1.05 $\pm$ 0.11 & 26.0 $\pm$ 7.0& 11\\
  KKH 98       & 4.4 $\pm$ 0.4 & $-$132.26 & 20.7 & 3.8 & 6.46 & 2.02 & 3.45 & 1.07 $\pm$ 0.12 & 46.0 $\pm$ 5.0&2\\
\hline\end{tabular}
\begin{center}
{References:
1-\cite{cote97}
2-\cite{hkk03}
3-\cite{hoffman96}
4-\cite{hr86}
5-\cite{kk04}
6-\cite{alfa}
7-\cite{alfa1}
8-\cite{matt95}
9-\cite{bouc07}
10-\cite{b08}
11-\cite{b06}
12-\cite{pustilnik}
13-\cite{bc04b}
14- Chengalur et al. 2008 (in preparation)
}
\end{center}
\end{table*}

%\subsection{Integrated HI emission}
%\label{ssec:HI}

       The global HI profiles for our sample galaxies, obtained from the coarsest resolution
data cubes (see Table~\ref{tab:obs}) are shown in Figure~\ref{fig:spec}. The  parameters derived 
from the global HI profiles for the whole  FIGGS sample are listed in Table~\ref{tab:hiresult}. 
The columns are as follows:
(1)~the galaxy name, 
(2)~the integrated HI flux along with the errorbars, 
(3)~the central heliocentric velocity  (V${\rm{_{sys}}}$), 
(4)~the velocity width at 50\% of the peak  ($\Delta V_{50}$), 
(5)~the HI diameter (in arcmin) at a column density of $\sim 10^{19}$ atoms cm$^{-2}$ (D${\rm{_{HI}}}$),
(6)~ the derived HI mass (M${\rm{_{HI}}}$), 
(7)~ the HI mass-to-light ratio (M$\rm{_{HI}/L_B}$), 
(8)~the ratio of the HI diameter to the Holmberg diameter. 
(9)~the ratio of the GMRT flux to the single dish flux  (FI/FI$_{\rm{SD}}$), 
(10)~the inclination as measured from the HI moment~0 maps (i${\rm{_{HI}}}$), and
(11)~the reference for the single dish fluxes.

As seen in Column(9) in Table~\ref{tab:hiresult}, the HI flux measured from the GMRT HI profiles for 
most FIGGS galaxies, in general, agree (within the errorbars) with the values obtained from the single 
dish observations. The average ratio of GMRT flux to single dish flux is $0.98$. This indicates that 
in general no flux  was missed because of the missing short spacings in our interferometric observations. 
However, for some galaxies the integrated flux derived from the GMRT observations is significantly
smaller than the single dish values. The GMRT fluxes could be lower than those obtained from  single 
dish measurements either because of (i)~a calibration error or (ii)~a large fraction of  HI being in
an extended distribution that is resolved out,   or (iii)~the single dish flux is erroneous, possibly 
because of confusion with galactic emission.  However, the flux of the point sources seen in the 
GMRT images are in good agreement with those listed in NVSS, indicating that our calibration is not 
at fault. 
We note that in the case of KKH~12, KKH~6, UGC~6456, UGCA~92 and KK~41 there is a strong local HI emission
at velocities very close to the systemic velocities, making it likely that the single dish  integrated 
flux measurements for these galaxies were contaminated by  blending of their HI emission
with that of the  galactic emission. In the case of KK~246, its HI spectrum was near the edge of the 
GMRT observing band, hence the flux could not be reliably estimated. 

The GMRT integrated HI emission of the sample galaxies, obtained from the coarsest resolution
data cubes (see Table~\ref{tab:obs}), overlayed on the optical Digitized Sky Survey (DSS)
images are shown in Figure~\ref{fig:mom0}.  

%\subsubsection{The HI Inclination}

\begin{figure}
\psfig{file=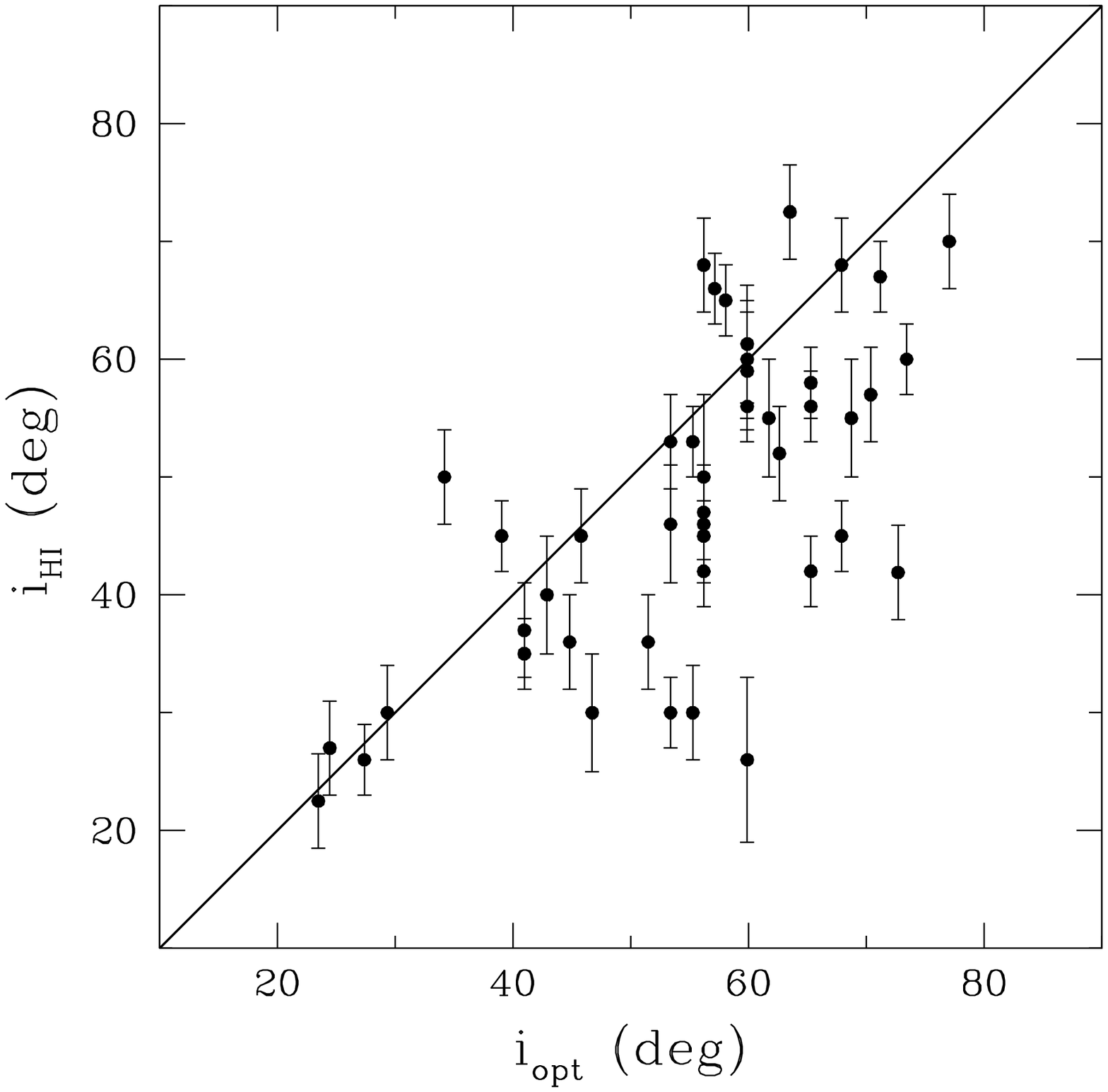,width=3.0truein}
\caption{A comparison of the morphological optical and HI inclination of the FIGGS sample.
The solid lines shows the case when the two inclinations are the same.
}
\label{fig:incl_figgs}
\end{figure}

The HI morphological inclinations (i${\rm{_{HI}}}$) for our sample galaxies were estimated 
from the integrated HI maps by fitting elliptical annuli to the HI images at various resolutions. 
For sample galaxies which have HI disks less extended than 2 synthesised beams (across the diameter 
of the galaxy) at the lowest HI resolution, could in principle be derived from the higher 
resolution HI maps. However, for most sample galaxies ellipse fitting to the high resolution
HI maps is not reliable because of clumpiness in the central high column density regions.
The derived inclination (without applying any correction for the intrinsic thickness of 
the HI disk)  is given in Column(10) in Table~\ref{tab:hiresult}. Figure~\ref{fig:incl_figgs} 
shows a comparison between the morphological inclination derived from the optical and HI isophotes
of the galaxy. No correction has been applied for the intrinsic thickness of the disk in both cases. 
The solid line shows a case when both inclinations are the same. We find that for 6 galaxies 
the HI inclination is significantly greater than the optical inclination (viz. KKH~11, UGC~6456, 
NGC~3741, UGC~8055, KK~230, KK~250). On the other hand, for many galaxies the optical inclination 
is found to be systematically higher than the inclination derived from the HI morphology. 
This result, if interpreted literally, suggest that the HI disks of these galaxies are 
thicker than the disks of their optical counterparts. However, we caution that a proper analysis
using deconvolved angular sizes of the the HI disks needs to be done before a firm
conclusion can be drawn.

%\subsubsection{The HI extent of low mass galaxies}

\begin{figure}
\psfig{file=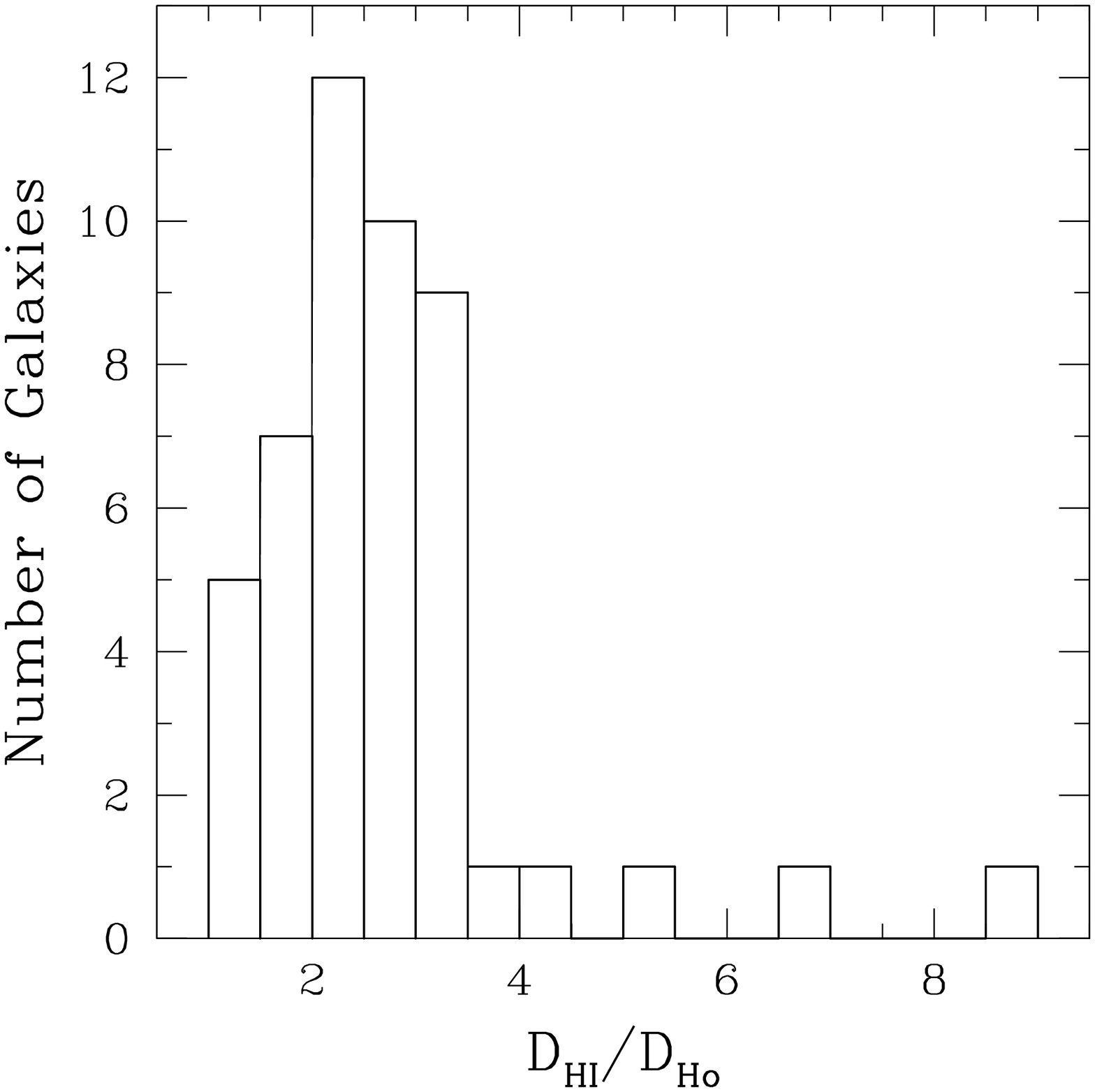,width=3.0truein}
\caption{The histogram of the extent of the HI disk (measured at N$_{HI}= 1 \times 10^{19}$ cm$^{-2}$),
normalised to the Holmberg diameter for the FIGGS sample.
}
\label{fig:hist_size}
\end{figure}

The diameter of the HI disk at a column density of ${\rm{N_{HI}\sim 10^{19}}}$ atoms cm$^{-2}$ 
(except for UGCA~92 where the  HI diameter is measured at ${\rm{N_{HI}\sim 10^{20}}}$ atoms cm$^{-2}$) 
estimated from the lowest resolution integrated HI emission maps is given in Column(7) of
Table~\ref{tab:hiresult}. The ratio of the HI diameter to the optical (Holmberg) diameter
for the sample is also given in Column(9) of the same table. Figure~\ref{fig:hist_size} shows 
the histogram of the derived HI extent of FIGGS at N${\rm{_{HI}\sim 10^{19}}}$cm$^{-2}$, normalised 
to the Holmberg diameter of the galaxy. The median HI extent of the FIGGS sample
(normalised to Holmberg diameter of the galaxy) is 2.4. For a comparison, Hunter (1997) 
using the data compiled from  the literature for comparatively bright Im galaxies found that
the ratio of D${\rm{_{HI}/D_{Ho}}}$ is somewhat smaller, viz. 1.5$-$2.  The extreme outliers
in Figure~\ref{fig:hist_size} is NGC~3741, our FIGGS data show it to have an HI extent 
of $\sim$ 8.3 times D${\rm{_{Ho}}}$ (Holmberg diameter). Follow-up WSRT+DRAO+GMRT observations 
resulted in HI being detected out to $\sim$8.8 D${\rm{_{Ho}}}$ $-$ NGC 3741 has the most 
extended HI disk known. For NGC~3741 the rotation curve could be derived out a record 
of $\sim$44 times the disk scale length and from the last measured point of the rotation 
curve we estimate the dynamical mass to light ratio,  M${\rm{_D/L_B}} \sim 149$ $-$
which makes it one of the ``darkest'' irregular galaxies known (Begum et al. 2005, 2008).

\begin{figure}
\psfig{file=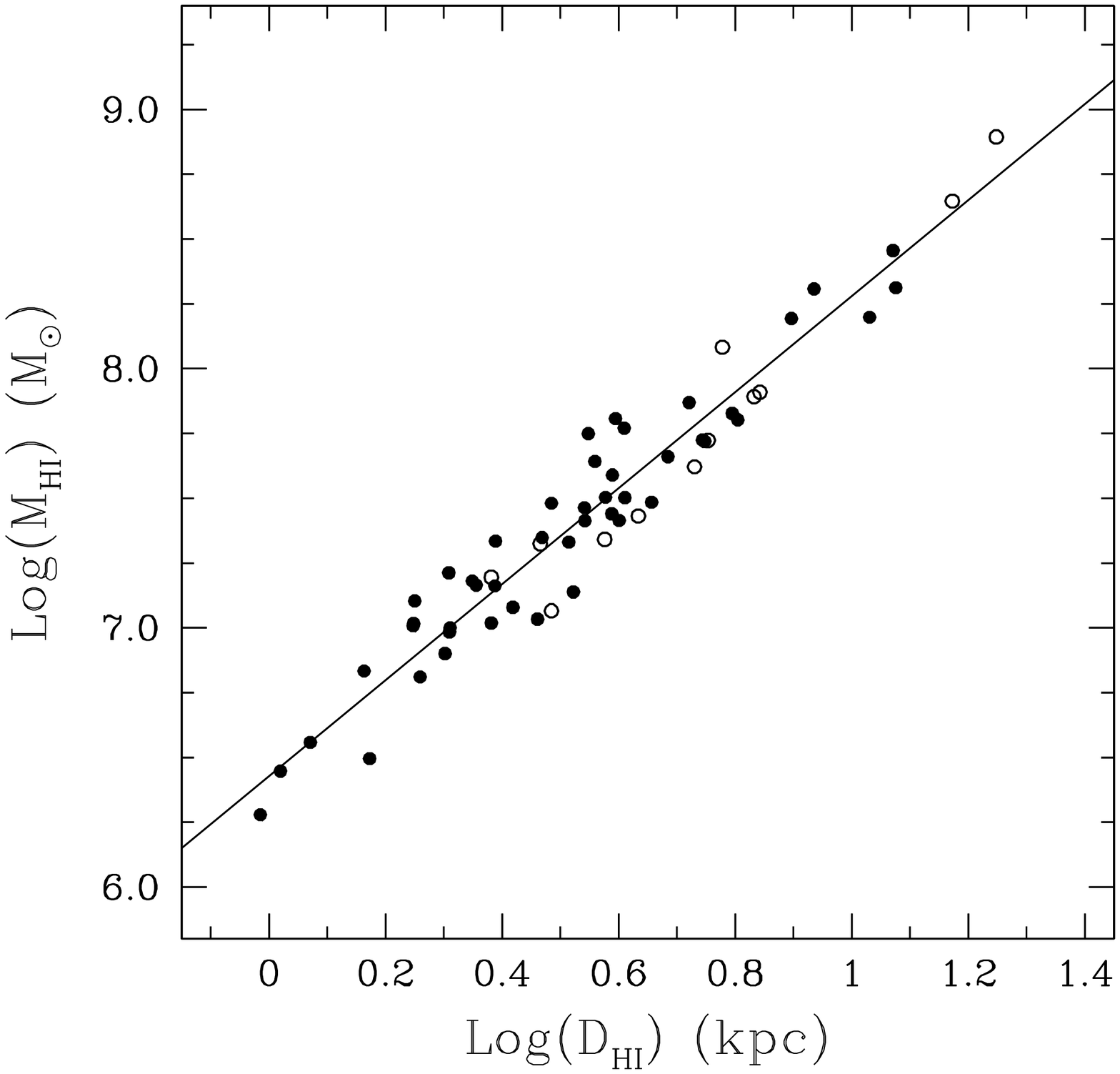,width=3.0truein}
\caption{ The HI mass for the FIGGS sample versus the HI diameter (measured at N${\rm{_{HI}\sim 10^{19}}}$ cm$^{-2}$). 
The solid line represents the fit to the data points.
Galaxies in FIGGS sample with TRGB distances are shown as solid points, while the remaining galaxies in the sample are
shown as open circles.
}
\label{fig:m_hi_size}
\end{figure}

Figure~\ref{fig:m_hi_size} shows a tight correlation between HI mass and the HI diameter,
measured at ${\rm{N_{HI}~of~ 1\times 10^{19}}}$ cm$^{-2}$ for FIGGS sample. The galaxies in 
FIGGS sample with accurate distances are shown as solid points, whereas the remaining galaxies 
are shown as open circle.  The best fit to the whole FIGGS sample shown as a solid line  gives

\begin{equation}
{\rm{log(M_{HI}) = (1.99\pm0.11) log(D_{HI}) + (6.08\pm0.06)}}
\label{eqn:mhi_dhi}
\end{equation}

\begin{figure}
\psfig{file=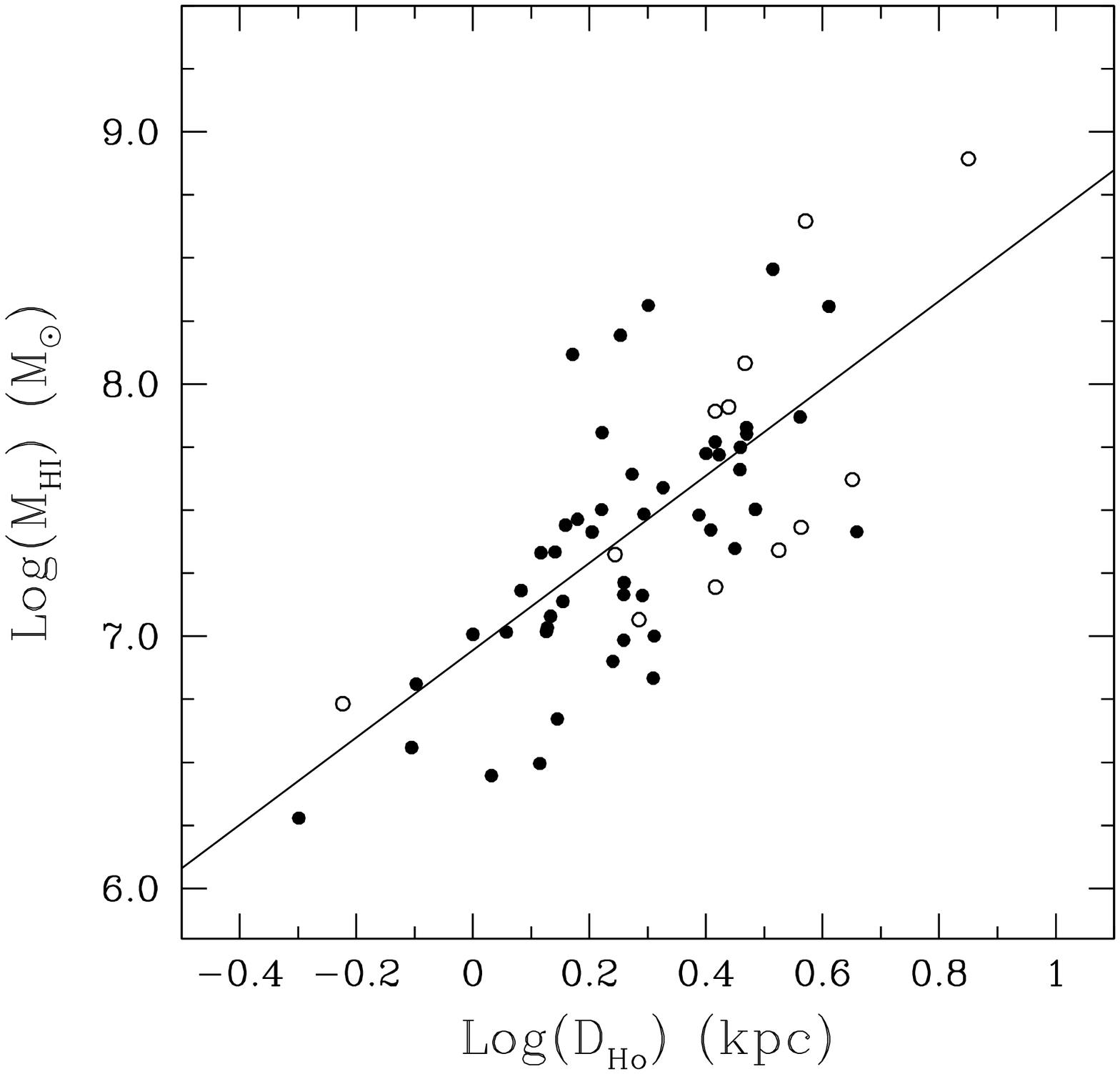,width=3.0truein}
\caption{ The HI mass for the FIGGS sample versus the Holmberg diameter. The solid line represents the fit to the data points.
Galaxies in FIGGS sample with TRGB distances are shown as solid points, while the remaining galaxies in the sample are
shown as open circles.
}
\label{fig:m_hi_optsize}
\end{figure}

{\hskip-0.75cm} The best fit relation was also derived using only the galaxies with TRGB distances,
however no significant difference was found between the best fit parameters derived in this case 
and that derived using the whole sample. Eqn.(\ref{eqn:mhi_dhi}) implies that the HI disks
of the FIGGS sample are well described as having an constant average surface mass density
$\sim 1.5$ M$_\odot$~pc$^{-2}$. A tight correlation between HI mass and the size of the HI 
disk has been noted earlier for spiral galaxies (e.g. \cite{br97}) and for brighter dwarf 
galaxies (Swaters 1999). For these samples the HI diameter was measured at a slightly
higher column density, viz. 1~M$_\odot$ pc$^{-2}$. For the FIGGS galaxies, the relationship
between the HI mass and the HI diameter measured at 1 M$_\odot$~pc$^{-2}$ is
${\rm{log(M_{HI}) = (1.96\pm0.10) log(D_{HI}) + (6.37\pm0.07)}}$, for comparison, \cite{br97}
%% Error bars for FIGGS computed assuming 10% errors in both MHI and DHI
measure ${\rm{log(M_{HI}) = (1.96\pm0.04) log(D_{HI}) + (6.52\pm0.06)}}$. The fit coefficients
overlap within the error bars. Hence from the FIGGS data we find that there is at best  marginal evidence for a decrease in average
HI surface density with decreasing HI mass; to a good approximation, the disks of gas rich 
galaxies, ranging over 3 orders of magnitudes in HI mass, can be described as being drawn from a family with
constant HI surface density. The HI mass also correlates with the optical (Holmberg) 
diameter (shown in  Fig.~\ref{fig:m_hi_optsize}), although with a larger scatter. A linear 
fit with a slope and intercept of 1.74$\pm$0.22 and 6.93$\pm$0.18, respectively is shown 
as a solid line. 
The larger scatter in the relation between M${\rm{_{HI}}}$ and the optical
diameter, also seen in sample of brighter dwarfs (e.g. \cite{swater99}), is probably indicative of a looser
coupling between the gas and star formation in dwarfs, compared to that in spiral galaxies.

\begin{figure}
\psfig{file=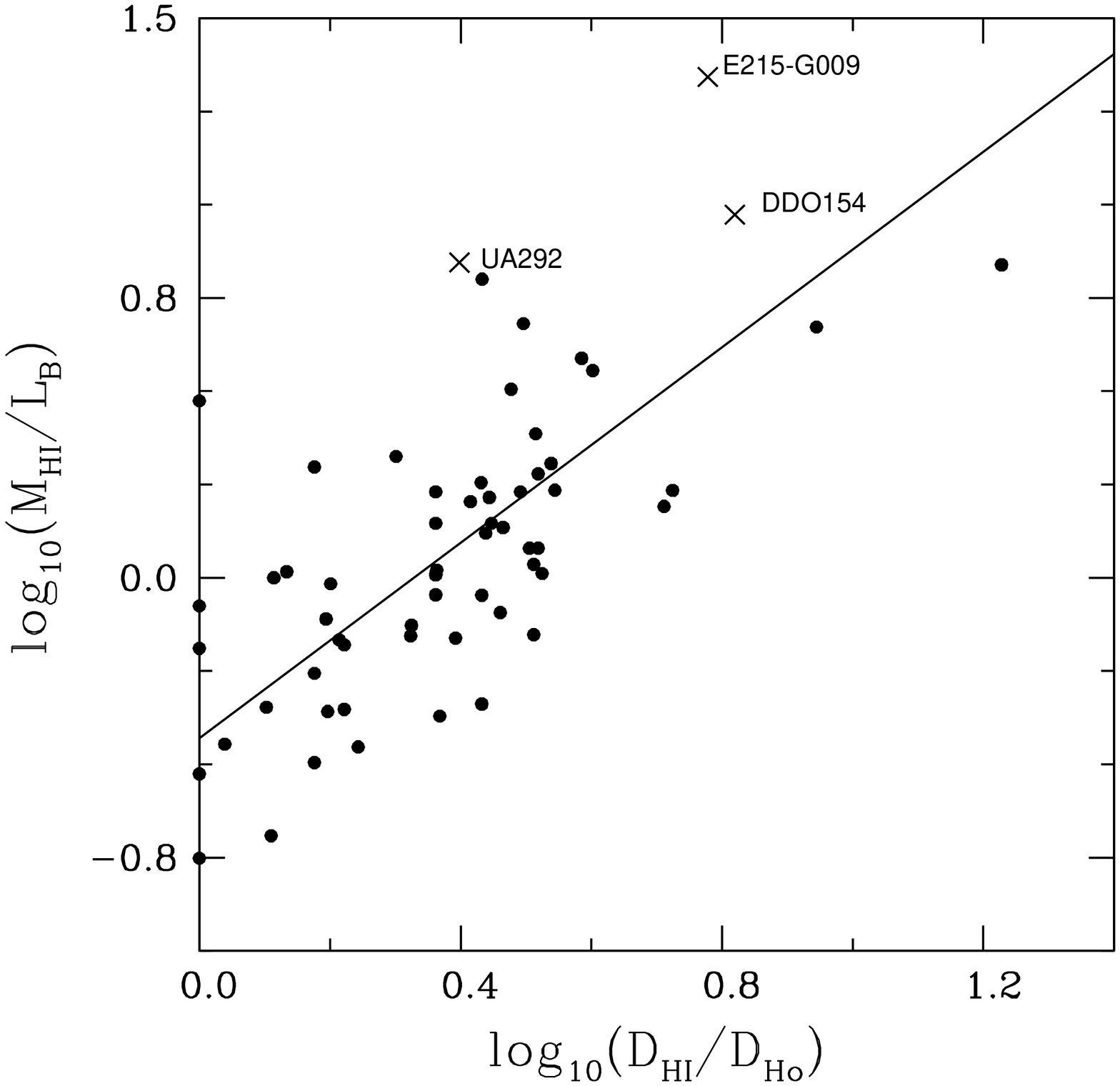,width=3.0truein}
\caption{ The log of HI mass to B band light ratio for the FIGGS sample plotted
as a function of the extent of the HI disk (measured at N${\rm{_{HI}\sim 1 \times 10^{19}}}$ cm$^{-2}$)
normalised to the Holmberg diameter. Crosses show three additional galaxies from the literature with high 
M${\rm{_{HI}/L_B}}$ and extended HI disks, UA292 (Young et al. 2003), ESO215-G?009 (Warren et al. 2004) and DDO 154 (
Carignan \& Purton 1998).
}
\label{fig:mtol_size}
\end{figure}

Figure~\ref{fig:mtol_size} shows the HI mass to light ratio for the FIGGS sample
plotted as a function of the HI extent, D${\rm{_{HI}/D_{Ho}}}$. A trend of an increase 
in the M${\rm{_{HI}/L_B}}$ with an increase in the HI extent of the galaxies is clearly seen. 
The best fit to the FIGGS sample shown as a solid line  gives 

\begin{equation}
{\rm{log(\frac{M_{HI}}{L_B}) = (1.31\pm0.18) log(\frac{D_{HI}}{D_{Ho}}) + (-0.43\pm0.08)}}
\end{equation} 

{\hskip-0.75cm} van Zee et al.(1995) from a HI mapping of a sample of low luminosity galaxies 
also found an evidence of an extended HI extent for high M${\rm{_{HI}/L_B}}$ galaxies.

\begin{figure}
\psfig{file=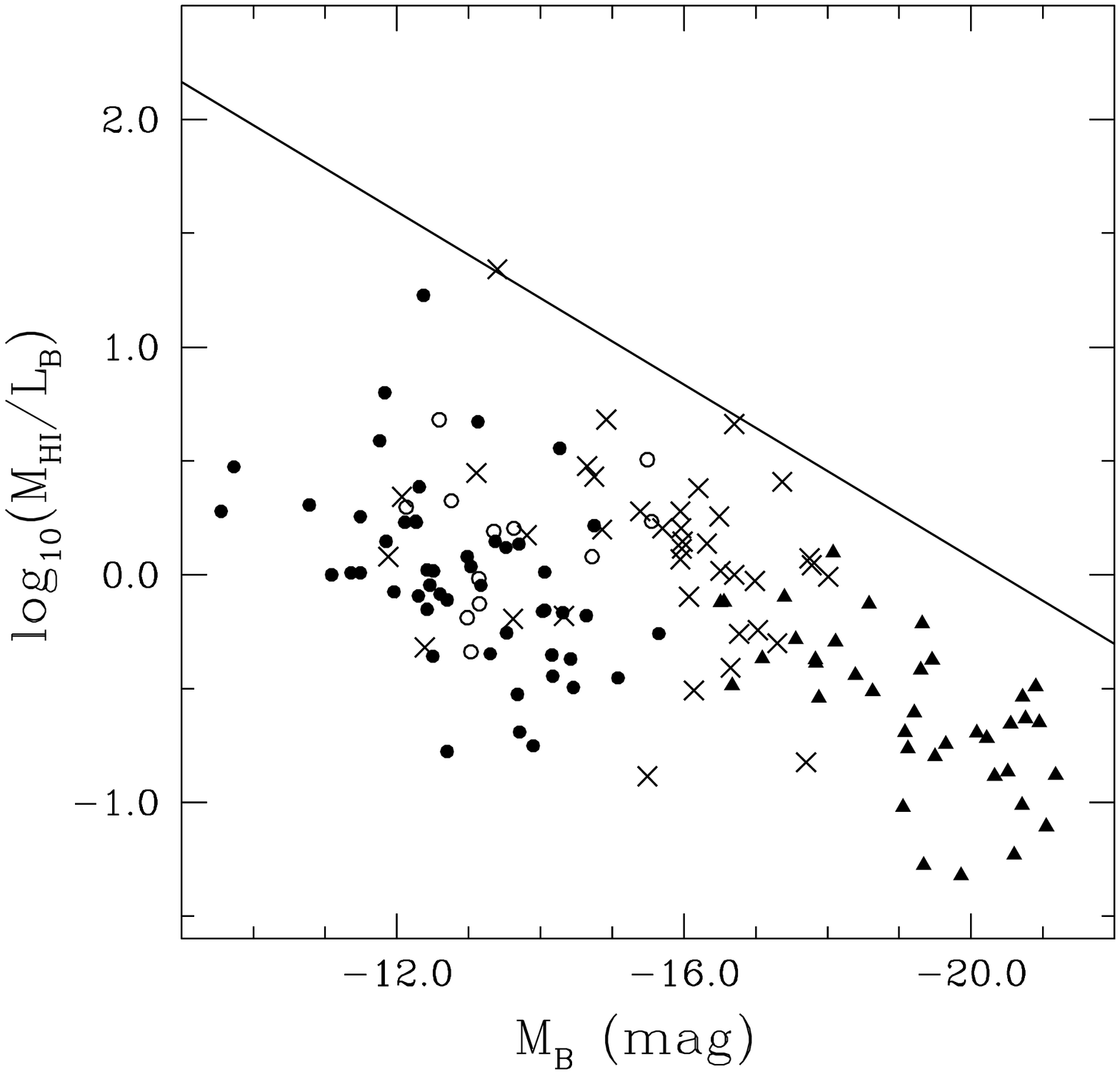,width=3.0truein}
\caption{The log of HI mass to light ratio vs. B band absolute magnitude. Galaxies from FIGGS sample
with TRGB distances are shown as solid points whereas the remaining FIGGS galaxies are shown as open circles.
Crosses are galaxies from Warren et al.(2007) and solid triangles from Verheijen (2001). The solid line
marks the locus of an upper envelope for the H I mass-to-light ratio at a given luminosity from Warren et al.(2007).
}
\label{fig:mtol_lb}
\end{figure}

\begin{figure}
\psfig{file=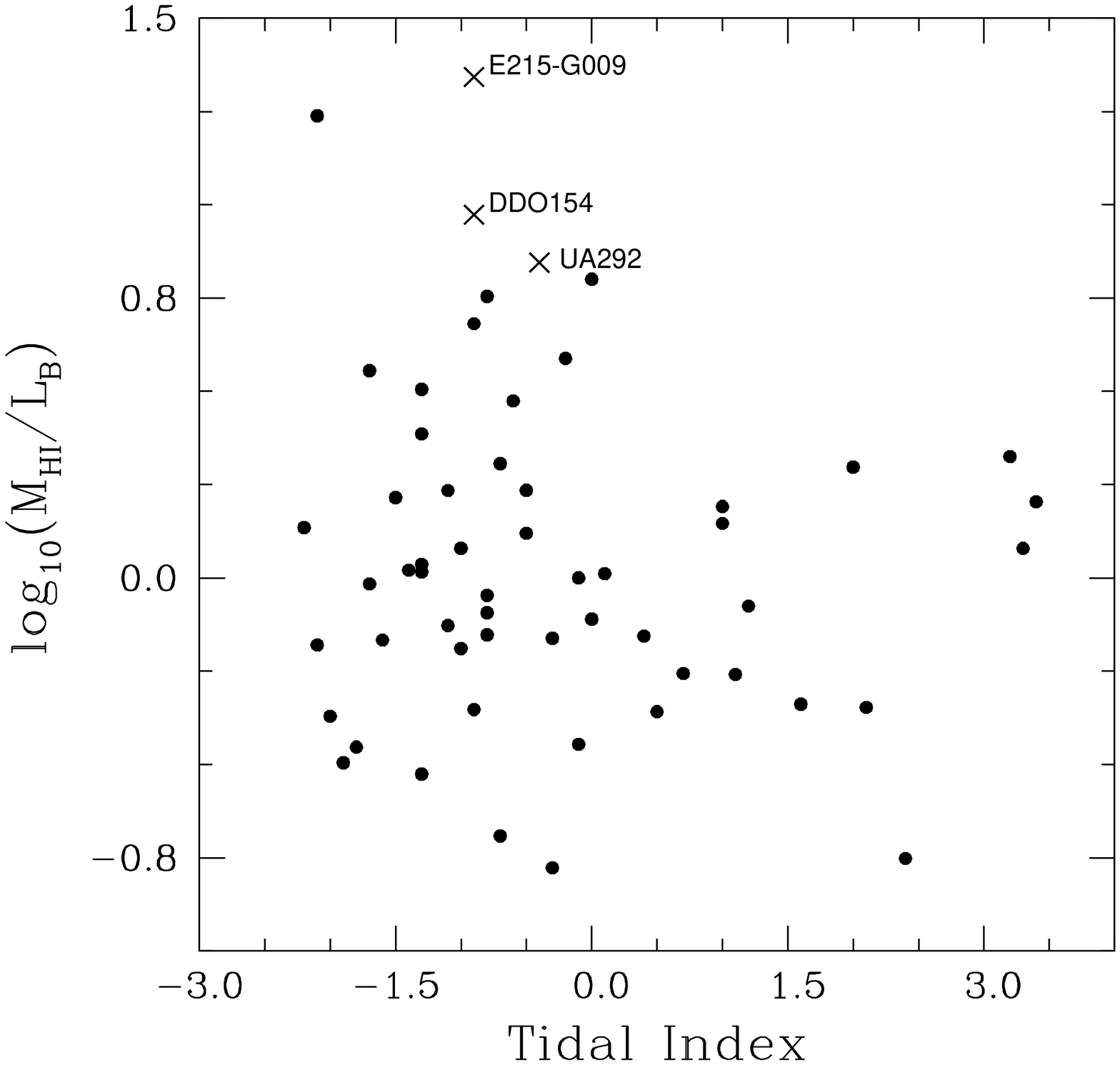,width=3.0truein}
\caption{ The log of HI mass to light ratio as a function of the tidal index for the FIGGS sample. 
Additional galaxies from literature with high M${\rm{_{HI}/L_B}}$ are also marked in the plot.
}
\label{fig:mtol_ti}
\end{figure}

\begin{figure}
\psfig{file=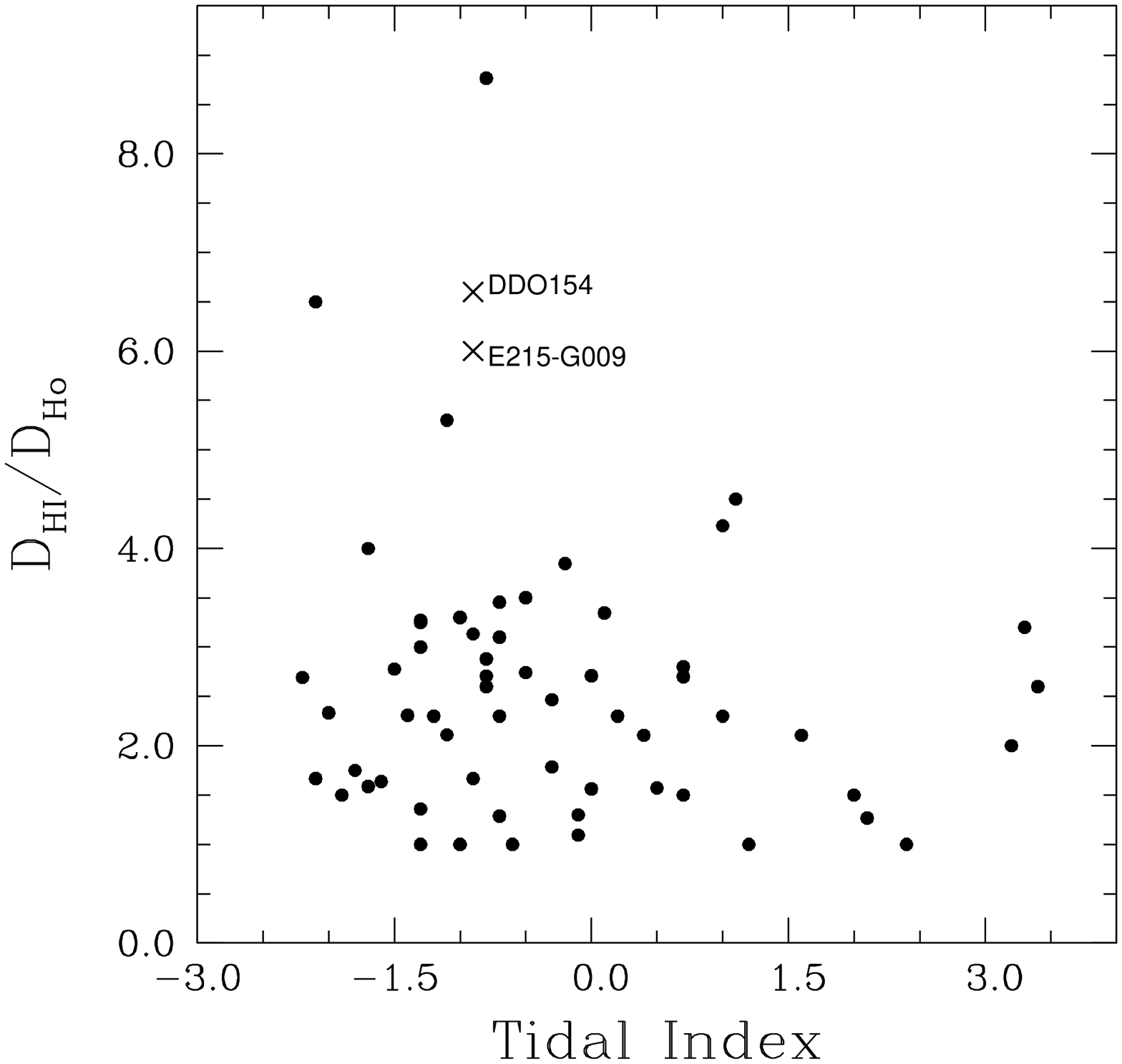,width=3.0truein}
\caption{ The HI extent of the FIGGS sample (normalised to the Holmberg radius) 
plotted as a function of the tidal index for the FIGGS sample. 
Additional galaxies from literature  with very extended HI disks viz. DDO 154 and ESO215-G?009
 and are also marked in the plot.
}
\label{fig:size_ti}
\end{figure}

%\subsubsection{The HI content of low mass galaxies}

Figure~\ref{fig:mtol_lb} shows  M${\rm{_{HI}/L_B}}$ for 
the FIGGS sample as a function of  M${\rm{_B}}$. The same quantity for several other spiral and 
dwarf galaxies, spanning a range in absolute B magnitude from M${\rm{_B \sim - 23}} $ to 
M${\rm{_B \sim -9}}$ is also plotted. The sample from which these galaxies have been drawn are listed 
in the figure caption. The galaxies in FIGGS sample with TRGB distances are shown as solid circles,
whereas the remaining FIGGS galaxies are shown as open circles. The solid line shows an empirically
determined upper envelope for M${\rm{_{HI}/L_B}}$ as a function of a M${\rm{_B}}$ from \cite{warren07}.
This upper envelope can be interpreted as a minimum fraction of the total baryonic mass which needs
to be converted into stars in order for a galaxy of a given baryonic mass to remain gravothermally 
stable (\cite{warren07}).  It is interesting to note that except for And~IV, all FIGGS galaxies lie 
much below this upper envelope. This implies that these galaxies have converted much more baryons 
into stars than the minimum required for remaining stable. In this context, it is interesting to 
note that the average gas fraction for the FIGGS sample is 0.7. Thus, for the majority of the 
dwarf galaxies in our sample, the baryonic mass is dominated by gas, rather than stars. 

In order to investigate the environmental dependence of the HI content for FIGGS galaxies, 
we plot  M${\rm{_{HI}/L_B}}$ for FIGGS sample as a function of tidal index (TI) 
(Figure~\ref{fig:mtol_ti}). Some additional gas rich galaxies with known HI extent are also 
plotted in the figure. TI is taken from \cite{kk04} and it represents the local mass density 
around a given galaxy, estimated using a large sample of galaxies within $\sim$ 10 Mpc of the 
Milky Way. A negative value of TI for a galaxy indicates that the galaxy is isolated, whereas 
a positive number indicates that the galaxy is in a dense environment. Figure~\ref{fig:mtol_ti}. 
shows that most of the FIGGS galaxies are in less dense environments, and 
that all the galaxies with high  M${\rm{_{HI}/L_B}}$ (i.e $\simgeq$ 2.5) 
have negative  tidal index i.e are isolated.  Figure~\ref{fig:size_ti} shows the HI extent of the
FIGGS sample, normalised to the optical (Holmberg) radius, plotted as a function of TI. 
As seen in the figure, the galaxies with very extended HI disks (${\rm{D_{HI}/D_{Ho}}}> 5.0$) are isolated.

To summarize, we have presented the first results from the Faint Irregular Galaxies GMRT Survey (FIGGS).
FIGGS is a large imaging program 
aimed at providing a comprehensive and statistically robust characterisation 
of the neutral ISM properties of extremely faint, nearby, gas rich, dIrr 
galaxies using the GMRT.  The GMRT HI data is supplemented with observations at other wavelengths.
The HI images in conjunction with the optical data will be used to investigate a variety
of scientific questions including the star formation feedback on the neutral ISM,
threshold for star formation, baryonic TF relation and dark matter distribution in low mass
galaxies. The optical properties of the FIGGS sample, GMRT observations 
and the main science drivers for the survey
 are described. The GMRT integrated
HI column density maps and the HI spectra for the sample galaxies are presented.  
The global HI properties of the FIGGS sample,  derived from the GMRT observations, and their 
comparison with the optical properties of the sample galaxies are also presented. 
%Using our FIGGS sample we confirmed the trend of increasing HI to optical diameter
%ratio with decreasing optical luminosity; the median ratio of D$_{\rm HI}$/D$_{\rm Ho}$ for
%the FIGGS sample is 2.4. Comparing our data with aperture synthesis surveys of bright spirals,
%we find  at best marginal evidence for a decrease in average surface density with decreasing
%HI mass; to a  good approximation for over 3 orders of magnitudes in HI mass the disks
%gas rich galaxies can be described as being drawn from a family with constant HI surface density.
A detailed comparison of the gas distribution, kinematics 
and star formation in the sample galaxies will be presented in the companion
papers.

\section*{Acknowledgments}

        The observations presented in this paper were made
with the Giant Metrewave Radio Telescope (GMRT). The GMRT is operated
by the National Center for Radio Astrophysics of the Tata Institute
of Fundamental Research.
Partial support for this work was provided by ILTP grant B-3.13.

\end{document}